\documentclass[%
 reprint,
unsortedaddress,
 amsmath,amssymb,
 aps,
prc,
]{revtex4-1}

\usepackage{color}
\usepackage{graphicx}
\usepackage{dcolumn}    
\usepackage{multirow}
\usepackage{bm}         
\usepackage{sidecap}
\usepackage{hyperref}   
\usepackage{xcolor}
\usepackage{amsmath}
\usepackage{color,colortbl}
\usepackage{booktabs} 
\usepackage{hhline}
\usepackage{ulem}
\usepackage{hyperref} 
\usepackage{physics}
\usepackage{footnote}

\definecolor{dark-red}{rgb}{0.,0.,0}
\definecolor{dark-blue}{rgb}{0.,0.,1}
\definecolor{medium-blue}{rgb}{0,0,1}
\definecolor{gray}{rgb}{0.85,0.85,0.85}

\hypersetup{
    colorlinks, linkcolor={dark-red},
    citecolor={dark-blue}, urlcolor={medium-blue}
}

\begin{document}

\title{Stellar electron capture rates based on finite temperature relativistic quasiparticle random-phase approximation}%

\author{A. Ravli\'c }%
\email{aravlic@phy.hr}
\affiliation {Department of Physics, Faculty of Science, University of Zagreb, Bijeni\v{c}ka c. 32, 10000 Zagreb, Croatia}%
\author{E. Y\"uksel}
\email{eyuksel@yildiz.edu.tr}
\affiliation{Yıldız Technical University, Faculty of Arts and Science, Department of Physics, Davutpasa Campus, TR-34220, Esenler/Istanbul, Turkey}
\author{Y. F. Niu}
\affiliation{School of Nuclear Science and Technology, Lanzhou University, Lanzhou, China.}
\author{G. Col\`o}
\affiliation{Dipartimento di Fisica, Universit\`a degli Studi di Milano, Milano, Italy}
\affiliation{INFN, Sezione di Milano, Via Celoria 16, 20133 Milano, Italy}
\author{E. Khan}
\affiliation{Institut de Physique Nucl\'eaire, Universit\'e Paris-Sud, IN2P3-CNRS, Universit\'e Paris-Saclay, F-91406 Orsay Cedex, France}
\author{N. Paar}
\email{npaar@phy.hr}
\affiliation{Department of Physics, Faculty of Science, University of Zagreb, Bijeni\v{c}ka c. 32, 10000 Zagreb, Croatia}%

\date{\today}%
\revised{February 2020}%

\begin{abstract}
The electron capture process plays an important role in the evolution of the core collapse of a massive star that precedes the supernova explosion. In this study, the electron capture on nuclei in stellar environment is described in the relativistic energy density functional framework, including both the finite temperature and nuclear pairing effects.
Relevant nuclear transitions $J^\pi = 0^\pm, 1^\pm, 2^\pm$ are calculated using the finite temperature proton-neutron quasiparticle random phase approximation with the density-dependent meson-exchange effective interaction DD-ME2.
The pairing and temperature effects are investigated in the Gamow-Teller transition strength as well as the electron capture cross sections and rates for ${}^{44}$Ti and ${}^{56}$Fe in stellar environment. 
It is found that the pairing correlations establish an additional unblocking mechanism similar to the finite temperature effects, that can allow otherwise blocked single-particle transitions.
Inclusion of pairing correlations at finite temperature can significantly alter the electron capture cross sections, even up to a factor of two for ${}^{44}$Ti, while for the same nucleus electron capture rates can increase by more than one order of magnitude. We conclude that for the complete description of electron capture on nuclei both pairing and temperature effects must be taken into account.
\end{abstract}

\maketitle

\section{Introduction}\label{sec:introduction}

Dynamics of core-collapse supernovae are determined by just two parameters: electron-to-baryon ratio $Y_e$ and the core entropy. These two parameters are mainly determined by the weak interaction processes in nuclei, in particular electron capture and $\beta$-decay \cite{BETHE1979487, JANKA200738, bethe_how}.
While the electron capture lowers the total number of available electrons in the stellar environment, and decreases $Y_e$, escaping neutrinos also decrease the core entropy. On the other hand, the $\beta$-decay acts in the opposite direction, and this process gains prominence for neutron-rich nuclei due to the increase in the available phase space \cite{JANKA200738}. The core of a massive star is stabilized by the electron degeneracy pressure until its mass does not exceed the Chandrasekhar mass $M_{ch} \sim Y_e^2$ \cite{BETHE1979487}. When the mass of the iron core reaches the Chandrasekhar mass $M_{ch}$, the electron degeneracy pressure can no longer hold the gravitational force and the core collapses. 

Due to their importance in the dynamics and evolution of massive stars,
different models have been employed to study the weak interaction processes in nuclei.
The first tabulation of weak interaction capture rates was presented by Fuller, Fowler, and Newman (FFN) \cite{fuller1980stellar, fuller1982stellar_part2, fuller1982stellar_part3,fuller1985stellar} using the independent particle model and nuclei with masses 21$<A<$60. It was shown that at higher temperatures, present in the presupernova collapse phase, nuclear weak interaction rates are dominated by Fermi and Gamow-Teller (GT) excitations. The first tables for the weak processes of $pf-$shell nuclei using large-scale shell-model calculations (LSSM) were also presented in Ref. \cite{LANGANKE20011} in the mass range  45$<A<$65. Later on, shell-model calculations with the GXPF1J interaction were used to evaluate electron capture rates \cite{PhysRevC.65.061301,PhysRevC.83.044619,Mori_2016}. Assuming the Nuclear Statistical Equilibrium (NSE) for the nuclear composition, the electron capture rates were calculated by a microscopic and hybrid approach for roughly 2700 nuclei \cite{JUODAGALVIS2010454}. Using the proton-neutron quasiparticle RPA (QRPA) with separable GT forces, stellar weak interaction rates were calculated for $fp-$ and $fpg-$ shell nuclei in Ref. \cite{NABI2004237}.

In the presupernova collapse, electron capture on $pf$-shell nuclei occurs at temperatures between 300 keV and 800 keV \cite{RevModPhys.75.819}. 
Therefore, the inclusion of temperature effect in the calculation of the relevant nuclear transitions and weak interaction processes of nuclei is quite essential at this stage of the collapse. In a previous microscopic study, the electron capture calculation on $fp-$shell nuclei was performed using the Shell-Model Monte Carlo (SMMC) approach at finite temperature \cite{PhysRevC.58.536}. Gamow-Teller strengths were used to calculate the electron capture cross sections and rates in zero-momentum transfer limits. While at low-temperatures (T$<$0.6 MeV) the changes in the GT strength are negligible, at higher temperatures the excited states are slightly shifted to lower excitation energies and the spectrum becomes broadened.
The temperature-unblocking effect was also studied in Ref. \cite{COOPERSTEIN1984591} for the GT transitions together with strength redistribution for the forbidden transitions with increasing temperature. Taking into account the first forbidden transitions in addition to the unblocked GT$^{+}$, it was shown that the electron capture on nuclei could dominate over the capture on free protons with
protons in the $pf$ shell and N $>$ 40. In Ref. \cite{PhysRevC.63.032801} a hybrid approach was employed: the SMMC was used to calculate finite temperature occupation numbers in the parent nucleus, and the excited states were obtained using the RPA. Using even-even germanium isotopes ${}^{68-76}$Ge, it was demonstrated that configuration mixing is strong enough to unblock the GT transitions at temperatures relevant for the core-collapse supernovae.

A fully self-consistent microscopic framework for the evaluation of nuclear weak-interaction rates at finite temperature based on Skyrme functionals was developed in Refs. \cite{refId0, PhysRevC.80.055801,PhysRevC.86.035805}. The single nucleon basis and corresponding thermal occupation factors were determined using the finite temperature Skyrme Hartree-Fock model. The relevant charge-exchange transitions were obtained for iron and germanium isotopes using the finite temperature RPA \cite{PhysRevC.80.055801}. Later on, the electron capture rates were also calculated for ${}^{54,56}$Fe and Ge isotopes \cite{PhysRevC.86.035805}. Within the relativistic energy density functional framework, the electron capture rates were also calculated using the finite temperature proton-neutron RRPA (FT-PNRRPA) for ${}^{54,56}$Fe and ${}^{76,78}$Ge \cite{PhysRevC.83.045807}. By increasing the temperature, it is found that the main peaks of the GT strength function are shifted towards lower excitation energies. Furthermore, additional peaks appear in the GT strength distribution that include transitions involving thermally unblocked single-particle levels. This modification of the GT strength has direct consequences on the electron capture cross sections and rates. It was also shown that electron capture becomes possible at lower energies of the incident electron and the effect of thermal unblocking plays an important role at high temperatures. Using the thermofield dynamics formalism, the electron capture rates were calculated in Ref. \cite{PhysRevC.81.015804}. Thermal evolution of GT${}^+$ strengths was presented for ${}^{54,56}$Fe and ${}^{76,78,80}$Ge. Recently, the thermal QRPA (TQRPA) approach was used to calculate the electron capture rates for N$=$50 nuclei and ${}^{56}$Fe \cite{PhysRevC.100.025801, PhysRevC.101.025805}. It was shown that thermal excitations can take significant contribution from the GT${}^+$ strength to electron capture rates. Also, the importance of forbidden transitions in electron capture calculations was demonstrated \cite{PhysRevC.100.025801, PhysRevC.101.025805}. The effect of the temperature on the spin-isospin response and beta decay rates of nuclei were also discussed using more advanced model which includes particle vibration coupling \cite{lit18}. However, the pairing effect was not taken into account and the calculations were limited to closed-shell nuclei. At present, there is no method to describe electron capture rates at finite temperature based on relativistic nuclear energy density functionals, with the pairing correlations taken into account.

In this work, we introduce the framework for the description of stellar electron capture cross sections and rates based on the relativistic energy density functionals~\cite{NIKSIC2011519, NIKSIC20141808},
using the finite temperature proton-neutron relativistic QRPA (FT-PNRQRPA)~\cite{yksel2019gamowteller, yksel2019nuclear}, that includes both the pairing and finite temperature effects. For the description of nuclear ground-state properties, we have used the finite temperature RMF theory combined with the Bardeen-Cooper-Schrieffer (BCS) approach. Unifying nuclear models for the description of the ground-state, nuclear transitions and electron capture cross sections, we have developed a consistent framework for the calculation of the electron capture rates on nuclei that are abundant in the core of presupernovae stars.

This paper is organized as follows. In Sec. \ref{sec:formalism} we describe the relativistic mean field theory and finite temperature Hartree BCS model. Then, formalisms for the FT-PNRQRPA and electron-capture cross sections and rates are introduced. In Sec. \ref{sec:results}, results are presented for the GT${}^+$ transition strength and electron capture cross sections and rates for ${}^{44}$Ti and ${}^{56}$Fe. Finally, conclusions and an outlook for future studies are given in Sec. \ref{sec:conclusion}.

\section{Formalism}\label{sec:formalism}

{Relativistic mean field (RMF) theory for finite nuclei is realized in the framework of relativistic nuclear energy density functionals~\cite{NIKSIC2011519, NIKSIC20141808}.}
Nucleons are treated as Dirac particles, and they can interact via meson exchange. In this study, we include isoscalar-scalar $\sigma$ meson, isoscalar-vector $\omega$ meson and isovector-vector $\rho$ meson, that build the minimal set of meson fields necessary to describe the bulk and single-particle nuclear properties.
The meson exchange model is described by the Lagrangian density \cite{RING1996193, GAMBHIR1990132, Reinhard_1989},

\begin{equation}
\mathcal{L} = \mathcal{L}_N + \mathcal{L}_m + \mathcal{L}_{int},
\end{equation}
$\mathcal{L}_N$ denotes free nucleon Lagrangian
\begin{equation}
\mathcal{L}_N = \bar{\psi}(i\gamma_\mu \partial^\mu - m )\psi,
\end{equation}
where $m$ is ``bare" mass of the nucleon and $\psi$ is Dirac spinor. Meson field Lagrangian is
\begin{align}
\begin{split}
\mathcal{L}_m &= \frac{1}{2}\partial_\mu \sigma \partial^\mu \sigma - \frac{1}{2}m_\sigma^2 \sigma^2 - \frac{1}{4}\Omega_{\mu \nu} \Omega^{\mu \nu} + \frac{1}{2}m_\omega^2 \omega_\mu \omega^\mu \\
&- \frac{1}{4} \vec{R}_{\mu \nu} \cdot \vec{R}^{\mu \nu} + \frac{1}{2} m_\rho^2 \vec{\rho}_\mu \cdot \vec{\rho}^\mu - \frac{1}{4} F_{\mu \nu} F^{\mu \nu},
\end{split}
\end{align}
with corresponding meson masses $m_\sigma, m_\omega, m_\rho$ and field tensors $\Omega_{\mu \nu}, \vec{R}_{\mu \nu}$ and $F_{\mu \nu}$
\begin{align}
\begin{split}
\Omega_{\mu \nu} &= \partial_\mu \omega_\nu - \partial_\nu \omega_\mu, \\
\vec{R}_{\mu \nu} &= \partial_\mu \vec{\rho}_\nu - \partial_\nu \vec{\rho}_\mu, \\
F_{\mu \nu} &= \partial_\mu A_\nu - \partial_\nu A_\mu,
\end{split}
\end{align}
corresponding to $\omega$-meson, $\rho$-meson and electromagnetic fields respectively. 
The vectors in isospin space are denoted with arrows above the symbols, while coordinate space vectors are boldfaced. The interaction Lagrangian is
given by,
\begin{equation}
\mathcal{L}_{int} = -g_\sigma \bar{\psi} \psi \sigma - g_\omega \bar{\psi} \gamma^\mu \psi \omega_\mu -g_\rho \bar{\psi} \vec{\tau} \gamma^\mu \psi \vec{\rho}_\mu - e \bar{\psi} \gamma^\mu \psi A_\mu,
\end{equation}
with coupling constants $g_\sigma, g_\omega, g_\rho$ and $e$. In this work, the density dependent meson-nucleon couplings are employed, and the DD-ME2 parameterisation is implemented in the calculations \cite{PhysRevC.66.024306, PhysRevC.71.024312}.
The energy density functional is given by
\begin{equation}
E_{RMF} = \int d^3 r \mathcal{H}(\boldsymbol{r}),
\end{equation}
where $\mathcal{H}(\boldsymbol{r})$ denotes Hamiltonian density. Since the meson-nucleon couplings are density-dependent, the rearrangement terms appear in the equation of motion, namely, they include derivatives of the couplings $g_\sigma, g_\omega$ and $g_\rho$ with respect to isovector density $\rho_v$. 
In this work, the ground-state properties of nuclei are calculated using the finite temperature Hartree BCS theory (FT-HBCS), assuming spherical symmetry \cite{GOODMAN198130, yuksel2014effect}. 
We should also mention that only the isovector pairing (T=1, S=0) contributes to the ground-state calculations and leads to the partial occupation of states. Within our current model, isoscalar pairing is not considered in the ground-state calculations because we do not consider proton-neutron mixing.
Using the FT-HBCS framework, occupation probabilities of single-particle states are given by
\begin{equation}
n_k = v_k^2(1-f_k) + u_k^2 f_k,
\end{equation}
where $v_k$ and $u_k$ are BCS amplitudes, and $f_k$ is the temperature dependent Fermi-Dirac distribution function
\begin{equation}
f_k = [1+ \text{exp}(E_k/k_B T)]^{-1},
\end{equation}
where $k_B$ and $T$ represent the Boltzmann constant and temperature, respectively. The $E_k$ is the quasiparticle (q.p.)  energy of a state and defined as $E_k = \sqrt{(\varepsilon_k-\lambda_q)^2 + \Delta_k^2}$ with $\varepsilon_k$ denoting single-particle energies and $\lambda_q$ chemical potentials, for either proton or neutron states. $\Delta_k$ represents the pairing gap of the given state. Central equation of the FT-HBCS theory is the gap equation, from which pairing gaps $\Delta_k$ are determined  \cite{ring2004nuclear}
\begin{equation}
\Delta_{k}= - \frac{1}{2} \sum_{k^\prime>0} v_{k \bar{k} k^\prime \bar{k}^\prime} \frac{\Delta_{k^\prime}\left(1-2 f_{k^\prime}\right)}{E_{k^\prime}},
\end{equation}
where $v_{k \bar{k} k^\prime \bar{k}^\prime}$ are matrix elements of pairing interaction $v_{k \bar{k} k^\prime \bar{k}^\prime} = \bra{k \bar{k}} V \ket{k^\prime \bar{k}^\prime}$ and $\bar{k}$ denotes the time-reversed single particle state $k$.  Further we denote only nonzero matrix elements with $v_{k \bar{k} k^\prime \bar{k}^\prime} = - G_{k k^\prime}$ and the gap equation is given by
\begin{equation}\label{eq:gap_equation}
\Delta_{k}=  \frac{1}{2} \sum_{k^\prime>0} G_{k k^\prime} \frac{\Delta_{k^\prime}\left(1-2 f_{k^\prime}\right)}{E_{k^\prime}}.
\end{equation}
In this work, we adopt monopole pairing force for which $G_{k k^\prime} = G\delta_{k k^\prime}$. The isovector pairing strengths are adjusted to reproduce the pairing gap values according to the three-point relation \cite{Bender2000, MOLLER199220}. Smooth cut-off weights are also introduced in the calculations to take into account the finite-range of pairing interaction \cite{Bender2000}. The cut-off weights are defined as
\begin{equation}
s_{k}=\frac{1}{1+\exp \left[\left(\varepsilon_{k}-\lambda_{q}-\Delta E_{q}\right) / \mu_{q}\right]},
\end{equation}
where $\mu_q=\Delta E_q/10$ and $\Delta E_q$ is fixed from the condition
\begin{equation}
\sum_{k} s_{k}=N_{q}+1.65 N_{q}^{2 / 3},
\end{equation}
where $N_q$ is total number of protons or neutrons. Now Eq. (\ref{eq:gap_equation}) reads
\begin{equation}\label{eq:gap_equation_with_cutoff}
\Delta_{k}=  \frac{1}{2} \sum_{k^\prime>0} G_{k k^\prime} \frac{s_{k^\prime}\Delta_{k^\prime}\left(1-2 f_{k^\prime}\right)}{ \sqrt{(\varepsilon_{k^\prime}-\lambda_q)^2 + \Delta_{k^\prime}^2 s_{k^\prime}^2}}.
\end{equation}
\\
\\

In the calculation of the excited states, both isovector (T = 1, S = 0) and isoscalar pairing (T = 0, S = 1) contribute to the FT-PNRQRPA residual interaction. While the isovector pairing is constrained
by the experimental data at the ground-state level, within the present framework the isoscalar
pairing contributes only in the residual interaction and can be constrained by the excitation properties~\cite{PhysRevC.69.054303}.
Following Ref.~\cite{PhysRevC.69.054303}, for the isoscalar pairing we employ a formulation
with a short
range repulsive Gaussian combined with a weaker longer
range attractive Gaussian
\begin{equation}\label{eq:isoscalar_pairing}
V_{12} = V_0^{is} \sum \limits_{j = 1}^2 g_j e^{-r_{12}^2/ \mu_j^2} \prod\limits_{S =1, T=0},
\end{equation}
where $\prod\limits_{S =1, T=0}$ denotes projector on T = 0, S = 1 states. 
For the ranges we use $\mu_1$ = 1.2 fm, and $\mu_2$ = 0.7 fm, and strengths are set to $g_1 =$ 1 and $g_2 = -$2~\cite{PhysRevC.69.054303}.
The residual isoscalar pairing strength $V_0^{is}$ is taken as a free parameter. Rather than constraining its value, in this work we study the effect of varying the isoscalar strength value on the excitations and electron capture cross sections and rates.
For the isovector pairing in the residual interaction, we employ the pairing part of the Gogny interaction \cite{PhysRevC.88.034308}. 
\\
\\
The FT-QRPA formalism was first developed in Ref. \cite{SOMMERMANN1983163}, however it was only applied to a schematic model. In order to study electron capture process at finite temperature, we employ the FT-PNRQRPA in change-exchange channel, introduced in Ref.~\cite{yksel2019gamowteller}. Here we only give a brief overview of the FT-PNRQRPA formalism, for more details see Ref.~\cite{yksel2019gamowteller}. The FT-PNRQRPA matrix is given by \cite{PhysRevC.96.024303, SOMMERMANN1983163}
\begin{equation}\label{eq:velika_matrica}
\begin{pmatrix}
\tilde{C} & \tilde{a} & \tilde{b} & \tilde{D} \\
\tilde{a}^\dag & \tilde{A} & \tilde{B} & \tilde{b}^T \\
-\tilde{b}^\dag & -\tilde{B}^* & -\tilde{A}^* & -\tilde{a}^T \\
- \tilde{D}^* & -\tilde{b}^* & -\tilde{a}^* & -\tilde{C}^*
\end{pmatrix} \begin{pmatrix}
\tilde{P} \\
\tilde{X} \\
\tilde{Y} \\
\tilde{Q} \\
\end{pmatrix} = E_\nu \begin{pmatrix}
\tilde{P} \\
\tilde{X} \\
\tilde{Y} \\
\tilde{Q} \\
\end{pmatrix},
\end{equation}
where $E_\nu$ represent the excitation energies, and eigenvectors $\tilde{P}, \tilde{X}, \tilde{Y}, \tilde{Q}$ are given by
\begin{align}
\tilde{X}_{ab} &= X_{ab}\sqrt{1-f_a-f_b}, \\
\tilde{Y}_{ab} &= Y_{ab}\sqrt{1-f_a-f_b}, \\
\tilde{P}_{ab} & = P_{ab}\sqrt{f_b-f_a}, \\
\tilde{Q}_{ab} &= Q_{ab}\sqrt{f_b-f_a},
\end{align}
and the matrix elements read \cite{PhysRevC.96.024303, SOMMERMANN1983163}
\begin{align}
\begin{split}
\label{Amatrix}\tilde{A}_{abcd} &= \sqrt{1-f_a-f_b} A^\prime_{a b c d} \sqrt{1-f_c-f_d} \\
 &+(E_a + E_b)\delta_{ac}\delta_{bd}, 
 \end{split}
 \end{align}
 \begin{align}\label{eq:sumbatrices}
\tilde{B}_{abcd} &= \sqrt{1-f_a -f_b} B_{abcd}\sqrt{1-f_c-f_d}, \\
\label{Cmatrix}\tilde{C}_{abcd} &= \sqrt{f_b - f_a}C^\prime_{abcd}\sqrt{f_d - f_c} \\
&+ (E_a - E_b)\delta_{ac}\delta_{bd},  \nonumber  \\
\tilde{D}_{abcd} &= \sqrt{f_b - f_a} D_{abcd} \sqrt{f_d - f_c}, \\
\tilde{a}_{abcd} &= \sqrt{f_b - f_a} a_{a b c d}\sqrt{1- f_c -f_d}, \\
\tilde{b}_{abcd} &= \sqrt{f_b - f_a} b_{abcd} \sqrt{1-f_c-f_d}, \\
\tilde{a}^+_{abcd} &= \tilde{a}^T_{abcd} =  \sqrt{f_d - f_c} a^+_{a b c d}\sqrt{1- f_a -f_b}, \\
\tilde{b}^T_{abcd} &= \tilde{b}^+_{abcd} =  \sqrt{f_d - f_c} b^+_{a b c d}\sqrt{1- f_a -f_b}.
\end{align}
Detailed expressions for matrix elements $\tilde{A} , \tilde{B}, \tilde{C},  \tilde{D}, \tilde{a}$ and $ \tilde{b}$ can be found in Refs. \cite{PhysRevC.96.024303, SOMMERMANN1983163}.
$E_{a(b)}$ denote proton(neutron) q.p. energies, and appear on diagonals of the submatrices $\tilde{A}$ and $\tilde{C}$ in Eq.  (\ref{eq:velika_matrica}). Submatrices $\tilde{A}$ and $\tilde{B}$ are non-vanishing at zero temperature and describe effects of the excitations of the q.p. pairs, while $\tilde{C}, \tilde{D}, \tilde{a}, \tilde{b}, \tilde{a}^+$ and $\tilde{b}^T$ start to gain importance with increasing temperature \cite{PhysRevC.96.024303}. 
The residual interaction in the particle-hole channel is obtained from the relativistic density-dependent meson-exchange effective interaction DD-ME2~\cite{PhysRevC.71.024312}, while in the particle-particle matrix elements the finite-range pairing interactions have been used in $T=1$ and $T=0$ channels.

The amplitude of a particular excitation with energy $E_\nu$ is given by \cite{PhysRevC.96.024303}
\begin{equation}\label{eq:amplitude}
A_{a b} = |\tilde{X}_{a b}^\nu|^2 - |\tilde{Y}_{a b}^\nu|^2 + |\tilde{P}_{a b}^\nu|^2 - |\tilde{Q}_{a b}^\nu|^2,
\end{equation}
and normalization condition is given by 
\begin{equation}
\sum \limits_{a > b} A_{a b} = 1.
\end{equation}

The reduced transition probability of an excited state is calculated using \cite{PhysRevC.96.024303} 
 \begin{align}\label{eq:gamow_teller_operator}
 \begin{split}
 B_\nu &= | \bra{\nu |} { \hat{F}_J }\ket{|\text{QRPA}}|^2 = \\
 & \left| \sum \limits_{cd} \left[ (\tilde{X}_{cd}^\nu + \tilde{Y}_{cd}^\nu)(v_c u_d + u_c v_d)\sqrt{1-f_c-f_d} \right. \right. \\
 &+ \left. \left. (\tilde{P}_{cd}^\nu + \tilde{Q}_{cd}^\nu)(u_c u_d-v_c v_d)\sqrt{f_d-f_c} \right] \bra{c |} { \hat{F}_J }\ket{ | d} \right|^2,
 \end{split}
 \end{align}
 
 where $\ket{\nu}$ is the excited state, $\hat{F}_J$ is the transition operator, and $\ket{\text{QRPA}}$ is the correlated FT-PNRQRPA vacuum state. For GT${}^\pm$ transitions, the operators are $\hat{F}_J = \sum_{i=1}^A \boldsymbol{\sigma} \tau_{\pm}$, where $\boldsymbol{\sigma}$ is Pauli spin matrix, and $\tau_\pm$ is isospin raising (lowering) operator. For the analysis of the GT${}^\pm$ states, the reduced transition probability reads
 \begin{eqnarray}\label{eq:bph}
 B_\nu &= & \left| \sum \limits_{c \geq d} b_{cd} \right|^2,
 \end{eqnarray}
where $b_{cd}$ corresponds to the partial contribution of a given particle-hole
 configuration to the transition probability.
 Numerical calculations have been performed with 20 oscillator shells in the ground-state and maximal energy $E_{cut} = 100$ MeV is used for the FT-PNRQRPA particle-hole configurations.
 \\
 \\
 The electron capture on nuclei is a weak interaction process,
 \begin{equation}
 e^- + {}^A_Z X_N \rightarrow {}^A_{Z-1} X_{N+1}^* + \nu_e.
 \end{equation}
In order to derive the electron capture cross section, we start from the Fermi golden rule
\begin{equation}\label{eq:fermijevo_pravilo}
\frac{d \sigma}{d \Omega} = \frac{1}{(2\pi)^2}\Omega^2 E_{\nu}^2 \frac{1}{2} \sum \limits_{lept. spin.} \frac{1}{2J_i +1} \sum \limits_{M_i, M_f} \left| \bra{f} \hat{H}_W \ket{i} \right|^2,
\end{equation}
where for the weak interaction part we are using current-current form of the Hamiltonian \cite{walecka1975muon, PhysRevC.6.719, walecka2004theoretical}
\begin{equation}\label{eq:struja-struja}
\hat{H}_W = - \frac{G}{\sqrt{2}}\int d^3 x j_{\mu}^{lept}(\boldsymbol{x})\hat{\mathcal{J}}_{\mu}(\boldsymbol{x}),
\end{equation}
where $G$ is the Fermi coupling constant, $j_\mu^{lept}(\boldsymbol{x})$ is lepton current and $\hat{\mathcal{J}}_\mu (\boldsymbol{x})$ is hadron current. 
\\
\\
Electron capture cross section can be obtained by multipole expansion of Eq. (\ref{eq:struja-struja}), performing the lepton traces using $\frac{\Omega^2}{2} \sum_{lept. spin}$ (where $\Omega$ is phase-space volume) and introducing some new operators to write expression in compact form (for details see Refs. \cite{walecka1975muon, PhysRevC.6.719, walecka2004theoretical}). The differential cross section is given by
\begin{widetext}
\begin{align}\label{eq:crosssection}
\begin{split}
\frac{d \sigma}{d \Omega} &= \frac{G^2 \cos^2\theta_c}{2\pi}\frac{F(Z, E_e)}{2J_i+1} \Biggl\{\sum\limits_{J \geq 1} \mathcal{W}(E_e, E_{\nu})\left\{[1-(\hat{\boldsymbol{\nu}}\cdot \hat{\boldsymbol{q}} )(\boldsymbol{\beta}\cdot \hat{\boldsymbol{q}})]\left[ |\bra{J_f}|\hat{\mathcal{T}}_{J}^{mag}| \ket{J_i}|^2 
+   |\bra{J_f}|\hat{\mathcal{T}}_{J}^{el}| \ket{J_i}|^2 \right]  \right.  \\
&\left. -2\hat{\boldsymbol{q}}\cdot( \hat{\boldsymbol{\nu}}-\boldsymbol{\beta}) \textrm{Re}\bra{J_f}|\hat{\mathcal{T}}_{J}^{mag}| \ket{J_i}\bra{J_f}|\hat{\mathcal{T}}_{J}^{el}| \ket{J_i}^* \right\} \\
&+ \sum\limits_{J \geq 0} \mathcal{W}(E_e, E_{\nu})\left\{\left[ 1-\hat{\boldsymbol{\nu}}\cdot\boldsymbol{\beta}+2(\hat{\boldsymbol{\nu}}\cdot \hat{\boldsymbol{q}} )(\boldsymbol{\beta}\cdot \hat{\boldsymbol{q}})\right] 
 |\bra{J_f}|\hat{\mathcal{L}}_{J}| \ket{J_i}|^2 + (1+\hat{\boldsymbol{\nu}}\cdot \boldsymbol{\beta})|\bra{J_f}|\hat{\mathcal{M}}_{J}| \ket{J_i}|^2 \right. \\
&-  \left. 2\hat{\boldsymbol{q}}\cdot(\hat{\boldsymbol{\nu}}+\boldsymbol{\beta})\textrm{Re}\bra{J_f}|\hat{\mathcal{L}}_{J}| \ket{J_i}\bra{J_f}|\hat{\mathcal{M}}_{J}| \ket{J_i}^* \right\}  \Biggr\}.
\end{split}
\end{align}
\end{widetext}
In here, $\boldsymbol{q} = \boldsymbol{\nu} - \boldsymbol{k}$ denotes the momentum difference between neutrino and electron, $\hat{\boldsymbol{\nu}}$ and $\hat{\boldsymbol{k}}$ are corresponding unit vectors. Electron velocity is $\boldsymbol{\beta} = \boldsymbol{k}/E_e$, where $E_e$ is the energy of incoming electron, $E_\nu$ is neutrino energy, and $\theta_c$ is the Cabbibo angle. The Fermi function $F(Z,E_e)$ takes into account distortion of electron wave function \cite{Kolbe_2003}. Nuclear recoil factor is
\begin{equation}
\mathcal{W}(E_e, E_{\nu}) = \frac{E_{\nu}^2}{1+E_e/M_T(1-\hat{\boldsymbol{\nu}}\cdot\boldsymbol{\beta)}},
\end{equation}
$M_T$ denotes mass of the target nuclei. Nuclear matrix elements between initial $\bra{J_i}$ and final $\ket{J_f}$ states, correspond to charge $\hat{\mathcal{M}}_J$, longitudinal $\hat{\mathcal{L}}_J$, transverse electric $\hat{\mathcal{T}}_J^{el}$ and transverse magnetic $\hat{\mathcal{T}}_J^{mag}$ multipole operators \cite{walecka1975muon, PhysRevC.6.719, walecka2004theoretical}. These matrix elements are calculated for selected total angular momentum and parity $J^\pi$, and in this work we calculate cross sections and capture rates for a number of multipoles. 

Energy of the outgoing neutrino is determined from the energy conservation
\begin{equation}\label{eq:kinematics}
E_\nu = E_e - E_{QRPA} - \Delta_{np} - (\lambda_n - \lambda_p),
\end{equation}
where $E_{QRPA}$ is the QRPA excitation energy, $\Delta_{np}$ is the mass difference between neutron and proton and $\lambda_{n(p)}$ is neutron (proton) chemical potential.
In the model calculations, the coupling constant $g_A$ in the axial-vector part of the transition operators~\cite{walecka1975muon,walecka2004theoretical}  is usually quenched from its free-nucleon value $g_A = -1.26$.
In the present study, we include the quenching of axial-vector coupling constant, using $g_A = -1.0$. This result is based on the RQRPA calculations of muon capture in Ref. \cite{PhysRevC.79.054323} that used quenching to reproduce the experimental data on muon capture rates.

Electron capture rates are calculated by \cite{PhysRevC.83.045807}
\begin{equation}\label{eq:rates}
\lambda_{ec} = \frac{1}{\pi^2 \hbar^3} \int \limits_{E_e^0}^{\infty} p_e E_e \sigma_{ec}(E_e)f(E_e, \mu_e, T)dE_e.
\end{equation}
In here, $E_e^0 = \text{max}(|E_{QRPA} + \Delta_{np} + (\lambda_n - \lambda_p)|, m_e c^2)$ is minimum electron energy for the capture process, and electron momentum is $p_e = \sqrt{E_e^2 - m_e^2 c^4}$.  The electron distribution is given by Fermi-Dirac distribution
\begin{equation}\label{eq:fermi-dirac}
f(E_e, \mu_e, T) = \frac{1}{\text{exp}\left(\frac{E_e-\mu_e}{kT}\right) +1},
\end{equation}
where $\mu_e$ is the chemical potential of the electrons, and $T$ is temperature. Chemical potential is determined by inverting the relation \cite{PhysRevC.83.045807}
\begin{equation}
\rho Y_e = \frac{1}{\pi^2 N_A} \left( \frac{m_e c}{\hbar} \right)^3 \int \limits_0^\infty (f_e - f_e^+)p^2 dp,
\end{equation}
where $\rho$ is baryon density, $Y_e$ is electron-to-baryon ratio, $N_A$ is Avogadro's number, and $f_e^+$ denotes Fermi-Dirac distribution of positrons, for which $\mu_e^+ = -\mu_e$. 

\section{Results}\label{sec:results}
It is known that the weak interaction process on $pf$-shell nuclei plays an essential role in the presupernovae evolution, which takes place at various stellar densities and temperatures \cite{PhysRevC.86.015809, Heger_2001}. Therefore, accurate determination of the GT${}^+$ strength  distribution under these conditions is important in the calculation of the electron capture cross sections and rates. In this section, we first present our results for the GT${}^+$ strength of ${}^{44}$Ti and ${}^{56}$Fe using the FT-PNRQRPA with DD-ME2 functional.
As mentioned above, the isovector pairing in the ground-state can be constrained by using
the experimental data on nuclei, whereas there is no clear consensus about the
strength of the isoscalar pairing in the QRPA residual interaction.
Therefore, both the residual isoscalar pairing and temperature are varied to study their influence on the GT${}^+$  strength distribution.
Afterward, their effects on the electron capture cross sections and rates are discussed, and the results are compared with other model calculations.

\subsection{\texorpdfstring{$^{44}$Ti} {} nucleus}

As the first case to test the framework introduced in Sec.~\ref{sec:formalism}, we consider $^{44}$Ti, as an open-shell nucleus with pronounced pairing effects, and explore its excitation properties, and EC cross sections and rates by varying the isoscalar pairing strength and temperature.
As mentioned in Sec. \ref{sec:formalism}, the ground-state properties are obtained using the finite temperature HBCS model. The isovector pairing strength is determined as $G_{n(p)}=41.4(26.5)$ MeV/A at zero temperature, according to the 3-point formula \cite{Bender2000}. Since we use the grand-canonical description in our model, the nucleus undergoes under a sharp phase transition at critical temperatures and pairing properties vanish. In this work, the critical temperature value for neutrons (protons) is obtained as $T_c=1.83 (0.88)$ MeV. 
Due to the large neutron pairing gap ($\Delta_n = 3.38$ MeV) in ${}^{44}$Ti, pairing effects are non-vanishing for relatively high temperatures (T$>$1.5 MeV). This temperature interval is also known as significant for core-collapse supernovae simulations \cite{JANKA200738}.

\begin{figure}[!ht]
  \begin{center}
\includegraphics[width=1\linewidth,clip=true]{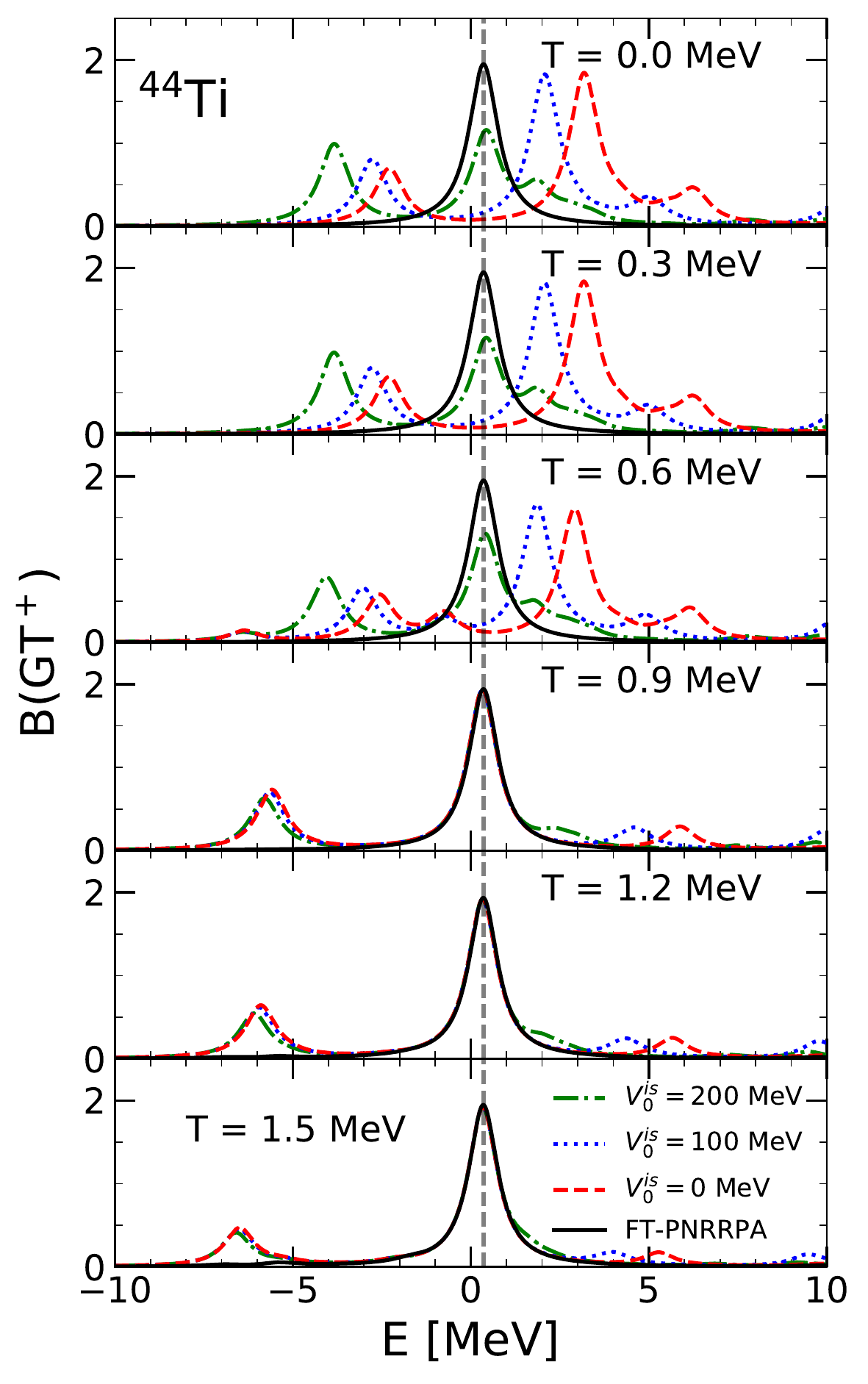}
  \end{center}
 \caption{The GT${}^+$ transition strength for ${}^{44}$Ti with respect to excitation energy of mother nucleus. The calculations are performed using the FT-PNRRPA (black line) and FT-PNRQRPA using various isoscalar pairing strength values $V_0^{is}$ = 0, 100 and 200 MeV at T = 0, 0.3, 0.6, 0.9, 1.2 and 1.5 MeV. Grey dashed line denotes the main FT-PNRRPA peak at T=0 MeV.} 
\label{fig:gtr_ti44}
\end{figure}

In Fig. \ref{fig:gtr_ti44}, the GT${}^+$ strength obtained with the FT-PNRRPA (black solid line) \cite{PhysRevC.83.045807} is shown together with the FT-PNRQRPA results for different values of the isoscalar pairing strength $V_0^{is}$. The calculations are performed at temperatures T=0, 0.3, 0.6, 0.9, 1.2 and 1.5 MeV. For demonstration purposes, the excited states are smoothed with a Lorentzian of 1 MeV width.
We start our analysis with the FT-PNRRPA results (without pairing correlations) at zero temperature (see Fig. \ref{fig:gtr_ti44}). The main peak is found at E = 0.36 MeV with the total strength B(GT${}^+$) = 3.06. This peak is mainly formed with the ($\pi 1f_{7/2},\nu1f_{5/2}$) transition. By increasing temperature up to T=1.5 MeV, the strength and excitation energy of the main peak almost do not change.

The FT-PNRQRPA results are also displayed in Fig. \ref{fig:gtr_ti44} at finite temperatures. As mentioned above, the proton-neutron isoscalar pairing is only included in the residual interaction part of the FT-PNRQRPA calculations, and it does not contribute to the ground-state calculations. Accordingly, the isoscalar pairing strength can be treated as a free parameter and we use various isoscalar pairing strength values ($V_0^{is} =0,100$ and 200 MeV) in our calculations to study its impact on the results. To simplify our discussion, we start our analysis with the results in case of no isoscalar pairing ($V_0^{is}=0$ MeV) at zero temperature (the topmost panel of Fig. \ref{fig:gtr_ti44}).
Using the FT-PNRQRPA, we found that the main peak is located at E = 3.18 MeV with B(GT${}^+$) = 2.83. Similar to the results using the FT-PNRRPA, the main contribution of this peak comes from ($\pi 1f_{7/2},\nu1f_{5/2}$) transition. It is known that the inclusion of the isovector pairing in the ground-state calculations unblocks the GT transitions, and leads to an increase in the quasiparticle energies of the states. Therefore, the main GT peak shifts to higher excitation energies compared to the FT-PNRRPA results. Apart from the main peak, two additional peaks appear with considerable strengths at E = -2.29 MeV (B(GT${}^+$) = 1.07) and 6.30 MeV (B(GT${}^+$) = 0.43). While the former one is mainly formed with $(\pi 1f_{7/2}, \nu 1f_{7/2})$, the strength of the latter mainly comes from $(\pi 2s_{1/2}, \nu 2s_{1/2})$ configuration. 

By increasing the isoscalar pairing strength at zero temperature, excited states start to shift towards lower energies due to the attractive nature of the residual isoscalar pairing.
While the main GT${}^+$ peak is obtained at E= 3.18 MeV without the isoscalar pairing, we obtain two peaks at E= 0.43 and 1.86 MeV with comparable strengths using the largest value of the isoscalar pairing strength. Furthermore, the strength of the main GT${}^+$ peak decreases, and the low-energy strength for E$<$0 MeV slightly increases. In Table \ref{tab:quasi_pairs_ti_qrpa}, the contribution from the most dominant configurations are presented for the main GT${}^+$ peak from model calculations with and without the isoscalar pairing.
For the two states, the table shows relative contributions of several configurations to the total norm (Eq. (\ref{eq:amplitude})) and their partial contributions to the transition strength $b_{cd}$  (see Eqs. (\ref{eq:gamow_teller_operator}) and (\ref{eq:bph})).
It is found that the ($l=l^\prime, j=j^\prime\pm1$) transitions also contribute to the low-energy strength as well as the ($l=l^\prime, j=j^\prime$) with the inclusion of the isoscalar pairing, which is consistent with the findings from previous studies \cite{PhysRevC.69.054303, PhysRevC.95.044301,PhysRevC.76.044307, PhysRevC.90.054335,BAI2013116, PhysRevC.60.014302,Sagawa_2016,NIU2018325}.
However, the contribution of the ($\pi 1f_{7/2},\nu1f_{5/2}$) transition decreases and the other transitions also contribute incoherently, which eventually leads to a slight decrease in the strength of this peak. Similar results are also obtained for the peaks at E$<$0 MeV, whereas the low-energy strength increases due to the coherent contribution of the transitions.
\begin{table*}
\caption{The quasiparticle configurations with the major contributions to the main GT${}^+$ states in ${}^{44}$Ti at zero temperature. 
The calculations are performed using the FT-PNRQRPA with and without isoscalar pairing.
The relative contribution of a particular transition to the total norm (see Eq. (\ref{eq:amplitude})) of an excited state and partial contributions to the strength for given configuration $b_{cd}$ obtained from calculation of GT${}^+$ matrix element (see Eqs. (\ref{eq:gamow_teller_operator}) and (\ref{eq:bph})) are shown.}\label{tab:quasi_pairs_ti_qrpa}
\begin{center}
\begin{tabular}{@{\extracolsep{4pt}}ccccccccc@{}}\hline\hline \\ [-1.5ex]
 &DD-ME2& \multicolumn{2}{c}{E=3.18 MeV $(V_0^{is} = 0$ MeV)} & &\multicolumn{2}{c}{E=0.43 MeV $(V_0^{is} = 200$ MeV)}  \\ \cline{3-4} \cline{5-7} \\[-1.ex] 
 &Configurations& Rel. strength (norm) (\%) & $b_{cd}$&  & Rel. strength (norm) (\%) & $b_{cd}$\\ [1.ex] 
\hline
 &  $(\pi 1f_{7/2}, \nu 1f_{5/2})$ & 97.86 & -1.77    &  & 67.91  & 1.50   \\
 &  $(\pi 1f_{5/2}, \nu 1f_{7/2})$ & 0.56  & 0.04     &  & 16.95  & -0.002  \\
 &  $(\pi 2p_{3/2}, \nu 2p_{3/2})$ & 0.44  & 0.03     &  & 3.46   & 0.06   \\
 &  $(\pi 1f_{7/2}, \nu 1f_{7/2})$ & 0.33  & -0.05    &  &1.75    & -0.26    \\
 &  $(\pi 1d_{3/2}, \nu 1d_{3/2})$ & 0.23  & 0.02     &  & 0.08    & -0.01   \\ 
 &  $(\pi 2p_{3/2}, \nu 2p_{1/2})$ & 0.09  & 0.01     &  & 3.20   & 0.06   \\ 
 &  $(\pi 2p_{1/2}, \nu 2p_{3/2})$ &       &          &  & 2.24   & 0.03    \\ 
\hline\hline
\end{tabular}
\end{center}
\vspace{-7mm}
\end{table*}

Below the critical temperature, one can observe significant differences between the FT-PNRQRPA and FT-PNRRPA strength
distributions, demonstrating important role of the pairing correlations. By increasing temperature from T=0 MeV toward T=0.6 MeV, the GT${}^+$ 
spectra show weak dependence on the temperature. One can observe only a slight decrease of the excitation energies and B(GT${}^+$) values. At higher temperatures, the paring effects weaken in the ground-state calculations and the FT-PNRRQPA residual interaction, and strength distributions are considerably different than at lower temperatures with the pairing interaction effects.

It is known that both the pairing and temperature can unblock the GT${}^+$ transitions. Below the critical temperatures, the unblocking effect of the temperature is not strong enough,  whereas the pairing correlations lead to the formation of new excited states in the low-energy part of spectra due to its unblocking effect on the quasi(single)-particle states.
With increasing temperature (below critical value), the isovector pairing effects also start to weaken.
The decrease in the isovector pairing effects has an impact on the ground-state properties of nuclei (i.e., occupation factors and single(quasi)-particle energies of states) and leads to a decrease in the quasiparticle energies of the states. Furthermore, the residual interaction, which contains both the particle-hole and isoscalar proton-neutron pairing interaction parts, weakens due to the temperature factors in front of the matrices (see Eqs. \ref{eq:sumbatrices}).

The evolution of the main GT${}^+$ peak for ${}^{44}$Ti with increasing temperature is shown in more details in Table \ref{tab:strengths_44ti} using the FT-PNRQRPA for $V_0^{is} = 200$ MeV. Up to T=0.9 MeV, the strength of the main peak increases
 and starts to shift to lower excitation energies. 
Compared to the results using the FT-PNRRPA, we obtain some part of the excitation spectrum with negative energies using the FT-PNRQRPA. 
Since those excitations are available to all electrons independent of their incident energy, they are going to have a considerable impact on electron capture calculations as we discuss below.

 \begin{table}
\centering
\caption{The evolution of the main GT${}^+$ peak energy and transition strength in ${}^{44}$Ti with temperature. The results are presented for the FT-PNRQRPA and $V_0^{is} = 200$ MeV is adopted.}\label{tab:strengths_44ti}
\begin{tabular}{ccc}
\hline
\hline\\ [-1ex]
&FT-PNRQRPA   &   \\
\hline\\ [-1ex]
T [MeV] & E [MeV] & B(GT${}^+$)  \\
\hline\\ [-1ex]
0.0 & 0.43 & 1.72 \\
0.3 & 0.43 & 1.73  \\
0.6 & 0.40 & 1.96  \\
0.9 & 0.32 & 3.01  \\
1.2 & 0.33 & 2.98  \\
1.5 & 0.34& 2.97  \\
\hline
\hline
\end{tabular}
\end{table}

\begin{figure*}[!ht]
 \begin{center}
\includegraphics[width=1\linewidth,clip=true]{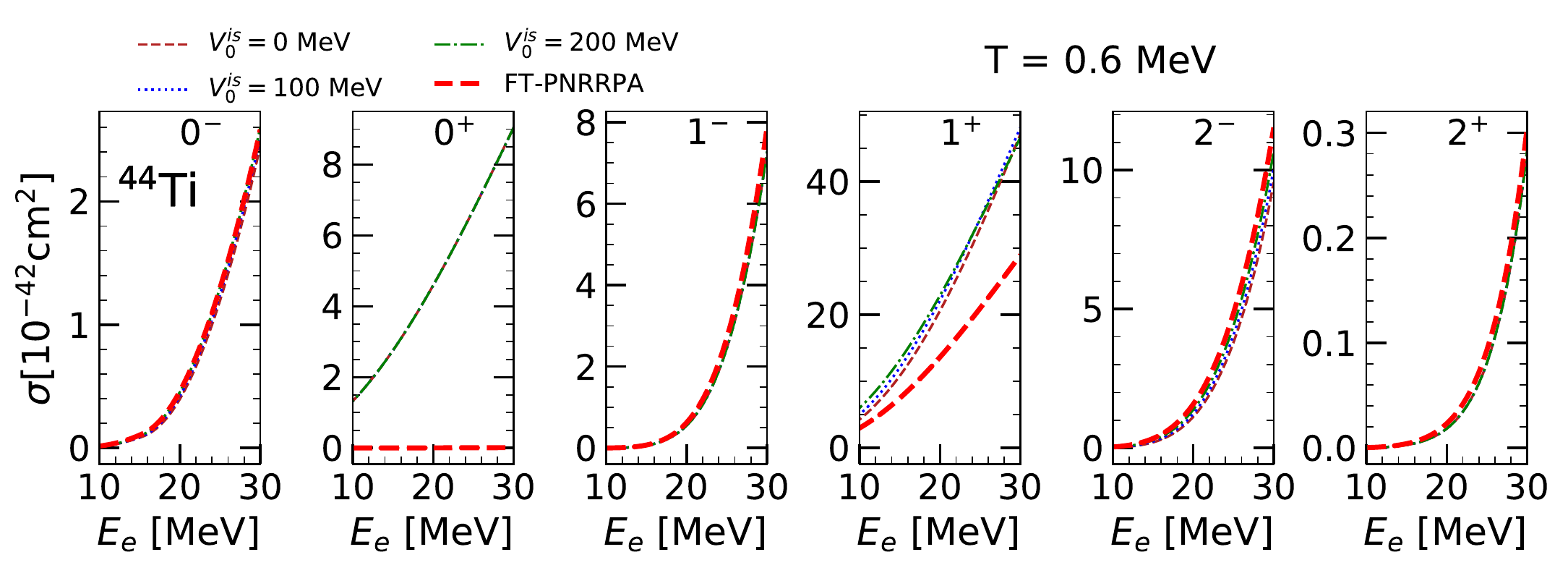}
\caption{Electron capture cross section for ${}^{44}$Ti decomposed into $J=0^\pm, 1^\pm, 2^\pm$ multipoles, shown with respect to the energy of incident electron $E_e$. The results are displayed for the FT-PNRRPA (red dashed line) and FT-PNRQRPA for different values of the isoscalar pairing strength $V_0^{is}$ at temperature T = 0.6 MeV.}\label{fig:cross_sec_multipoles_ti44}
  \end{center}
\end{figure*}

In Fig. \ref{fig:cross_sec_multipoles_ti44}, we present for ${}^{44}$Ti the EC cross sections for $J=0^\pm, 1^\pm, 2^\pm$ multipoles at T=0.6 MeV. The calculations are performed using both the FT-PNRRPA and FT-PNRQRPA, and various isoscalar pairing strength values are used in the latter case to study its impact on the results. Since natural parity transitions are determined, to a good approximation, only by the isovector pairing (pairing part of Gogny interaction, cf. Sec. \ref{sec:formalism}) in residual interaction, the $J= 0^+, 1^-, 2^+$ transitions show no dependence on varying the isoscalar pairing strength. On the other hand, the isoscalar pairing is present only for unnatural parity transitions ($0^-, 1^+, 2^-$) in the residual interaction.
Using the FT-PNRQRPA, the EC cross section takes contributions from all multipoles $J=0^\pm, 1^\pm, 2^\pm$, where $1^+$ has the most significant contribution and $2^+$ multipole has the lowest contribution. 
It is also seen that $J= 0^-, 2^-$ multipoles display a mild dependence on the changes in the isoscalar pairing strength. Compared to the FT-PNRRPA results, the calculations using the FT-PNRQRPA show that the $1^+$ transition has larger impact on the cross section due to the pairing effects. As mentioned above, the FT-PNRQRPA predicts considerable amount of $1^+$ excitation strength at negative excitation energies (see Fig. \ref{fig:gtr_ti44}).
The strength of the low-energy peak is found to be slightly higher for the larger
values of the isoscalar pairing strength. Therefore, the calculated EC cross sections
at lower electron energies increase with the isoscalar pairing strength.

At higher energies of incident electron, broader range of GT${}^+$ strength can be excited by incoming electrons, hence results are very similar for different values of $V_0^{is}$. At incident electron energy of 30 MeV, the FT-PNRQRPA gives for $1^+$ multipole larger cross section value compared to the FT-PNRRPA by a factor of 2, irrespective of the isoscalar pairing strength. This demonstrates the importance of including pairing correlations in EC calculations. Results for other multipoles display only slight deviations between the FT-PNRRPA and FT-PNRQRPA calculations, except for $0^+$ case.
From Fig. \ref{fig:cross_sec_multipoles_ti44} it is seen that the EC cross section for $0^+$ multipole is considerably increased using the FT-PNRQRPA. 
To explain this result it would be useful to investigate isobaric analog state   (IAS${}^+$). At T=0.6 MeV, the FT-PNRRPA predicts almost no IAS${}^+$ excitations, hence its contribution to the total EC cross section is negligible. On the other hand, the FT-PNRQRPA calculations predict a strong low-energy peak at E = -3.70 MeV with strength B(IAS${}^+$) = 0.63 and giving rise to higher $0^+$ EC cross sections. The main contribution for this low-energy peak comes from the $(\pi 1f_{7/2}, \nu 1f_{7/2})$ transition.

\begin{figure}[!ht]
  \begin{center}
\includegraphics[width=1\linewidth,clip=true]{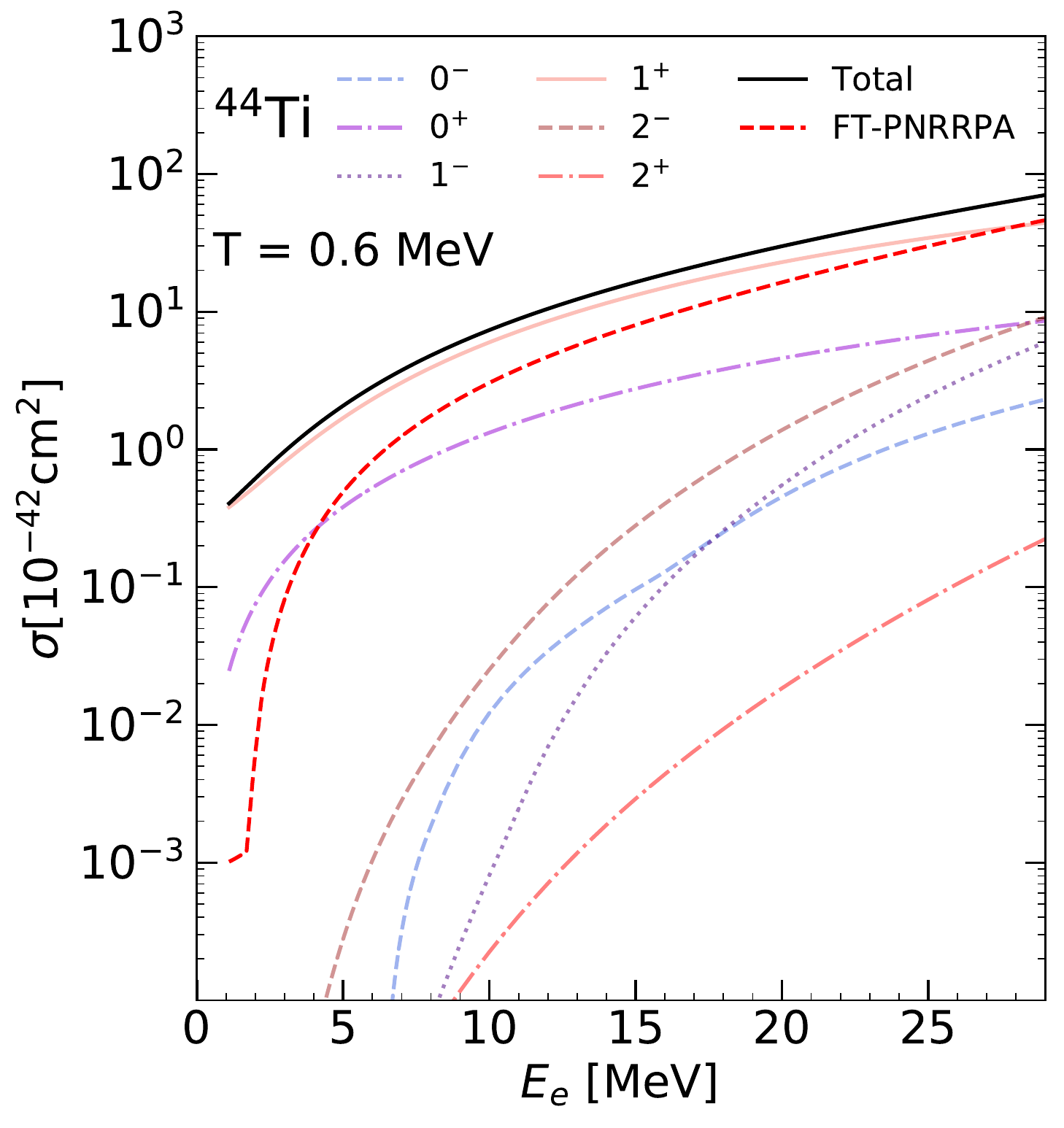}
\caption{Electron capture cross section for ${}^{44}$Ti with respect to the energy of incident electron $E_e$ for $J=0^\pm, 1^\pm, 2^\pm$ multipoles at temperature T = 0.6 MeV. Total FT-PNRQRPA cross section (black solid line) and FT-PNRRPA cross section (red dashed line) are plotted on the same figure. The isoscalar pairing strength is taken as $V_0^{is} = 200$ MeV.}\label{fig:cross_section_T3TI56.png}
  \end{center}
\end{figure}

Figure \ref{fig:cross_section_T3TI56.png} shows the total EC cross sections for ${}^{44}$Ti with contributions from all multipoles as well as the contributions from
each channel using the FT-PNRQRPA (black full line) and FT-PNRRPA (red dashed line). For demonstration purposes, the FT-PNRQRPA results are displayed for  $V_0^{is}=200$ MeV at T=0.6 MeV. 
The FT-PNRQRPA calculations predict larger cross sections compared to the FT-PNRRPA, as explained in previous analysis of multipole contributions shown in Fig.~\ref{fig:cross_sec_multipoles_ti44}. 
We also find that the largest contributions to the EC cross section at low energies of incoming electron come from the $1^+$ and $0^+$ transitions, respectively.
At higher energies of incident electron ($E_e \sim 30$ MeV), $1^-$ and $2^-$ transitions also have sizeable contributions to the total EC cross section. This result confirms that for increasing electron energies forbidden multipoles also become non-negligible, even for a light nucleus like ${}^{44}$Ti.

\begin{figure}[!ht]
  \begin{center}
\includegraphics[width=1\linewidth,clip=true]{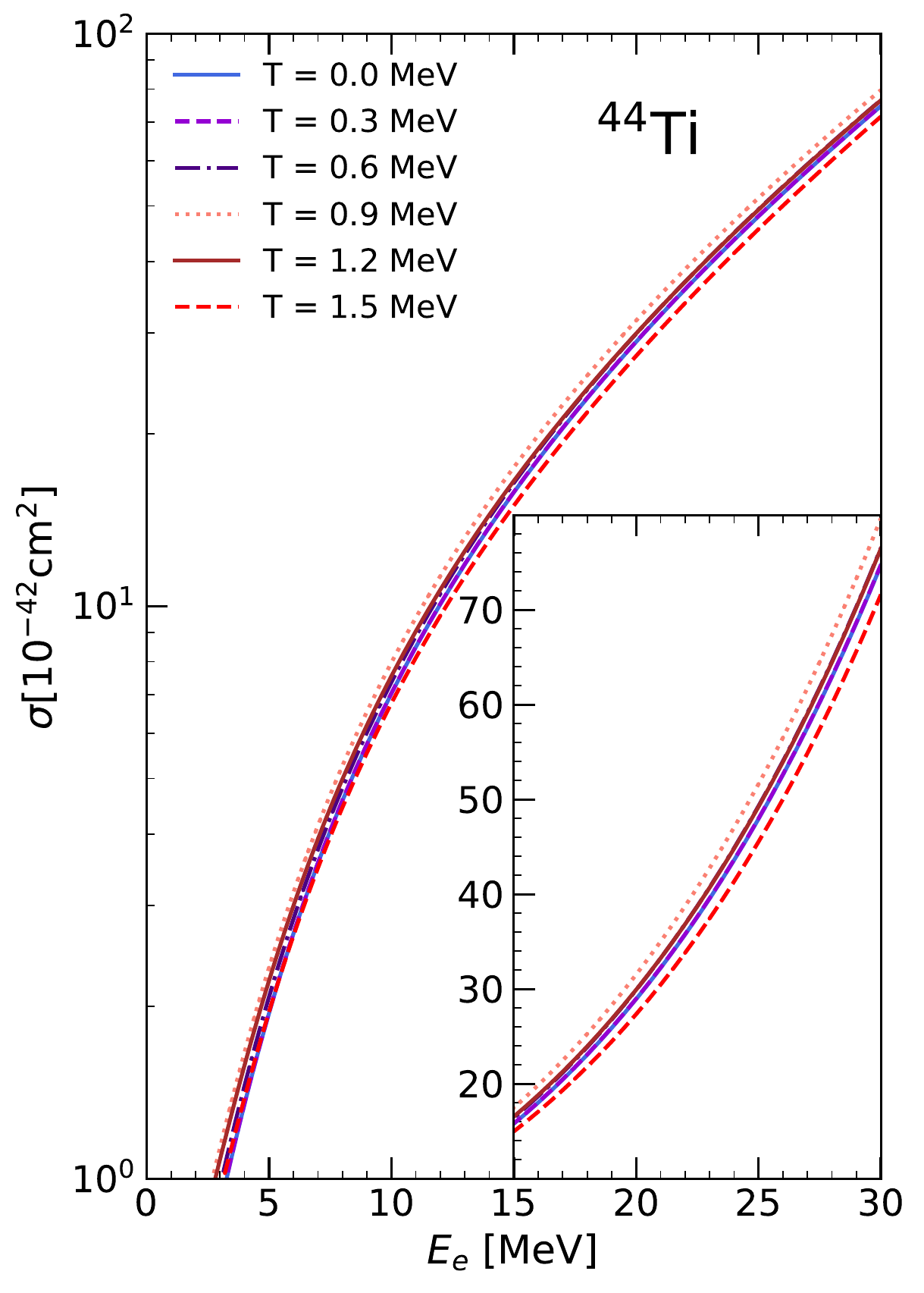}
\caption{Electron capture cross sections for ${}^{44}$Ti with respect to energy of incident electron at T = 0, 0.3, 0.6, 0.9, 1.2 and 1.5 MeV. The calculations are performed using the FT-PNRQRPA with   $V_0^{is} = 200$ MeV. In the insertion, the electron capture cross sections are displayed for a particular range of incident electron energy.}\label{fig:temp_ti44.png}
  \end{center}
\end{figure}

In Fig. \ref{fig:temp_ti44.png}, the EC cross sections for ${}^{44}$Ti are presented at finite temperatures.
The calculations are performed using the FT-PNRQRPA and the isoscalar pairing strength is fixed to $V_0^{is}=200$ MeV.
For temperatures up to T=0.9 MeV, the EC cross section closely follows temperature dependence of the main GT${}^+$ peak from Table \ref{tab:strengths_44ti}.
The overall trend is increasing the strength in the main peak with increasing temperature, as already discussed. 
The EC cross section values increase considerably at T=0.9 MeV due to the increase in the strength of the main peak as shown in Tab. \ref{tab:strengths_44ti}.
By further increasing temperature to T=1.2 MeV and T=1.5 MeV, it is seen that the EC cross sections decrease. 
From Table \ref{tab:strengths_44ti}, it can be seen that this downward shift in cross sections cannot be explained just by considering the main GT${}^+$ peak. The overall GT${}^+$ strength also decreases with temperature for T$>$0.9 MeV due to the weakening of pairing effects. At T=1.2 MeV, $\sum$B(GT${}^+$) = 4.56, while at T=1.5 MeV, $\sum$B(GT${}^+$) = 4.29, which explains the lowering of EC cross sections. Similar results are also obtained for $V_0^{is} = 0$ and $V_0^{is} = 100$ MeV. Analysis of the EC cross sections with multipole contributions by varying the isoscalar pairing and temperature is also important to explain the behavior of the EC rates, as we discuss below.

\begin{figure}[!ht]
  \begin{center}
\includegraphics[width=1\linewidth,clip=true]{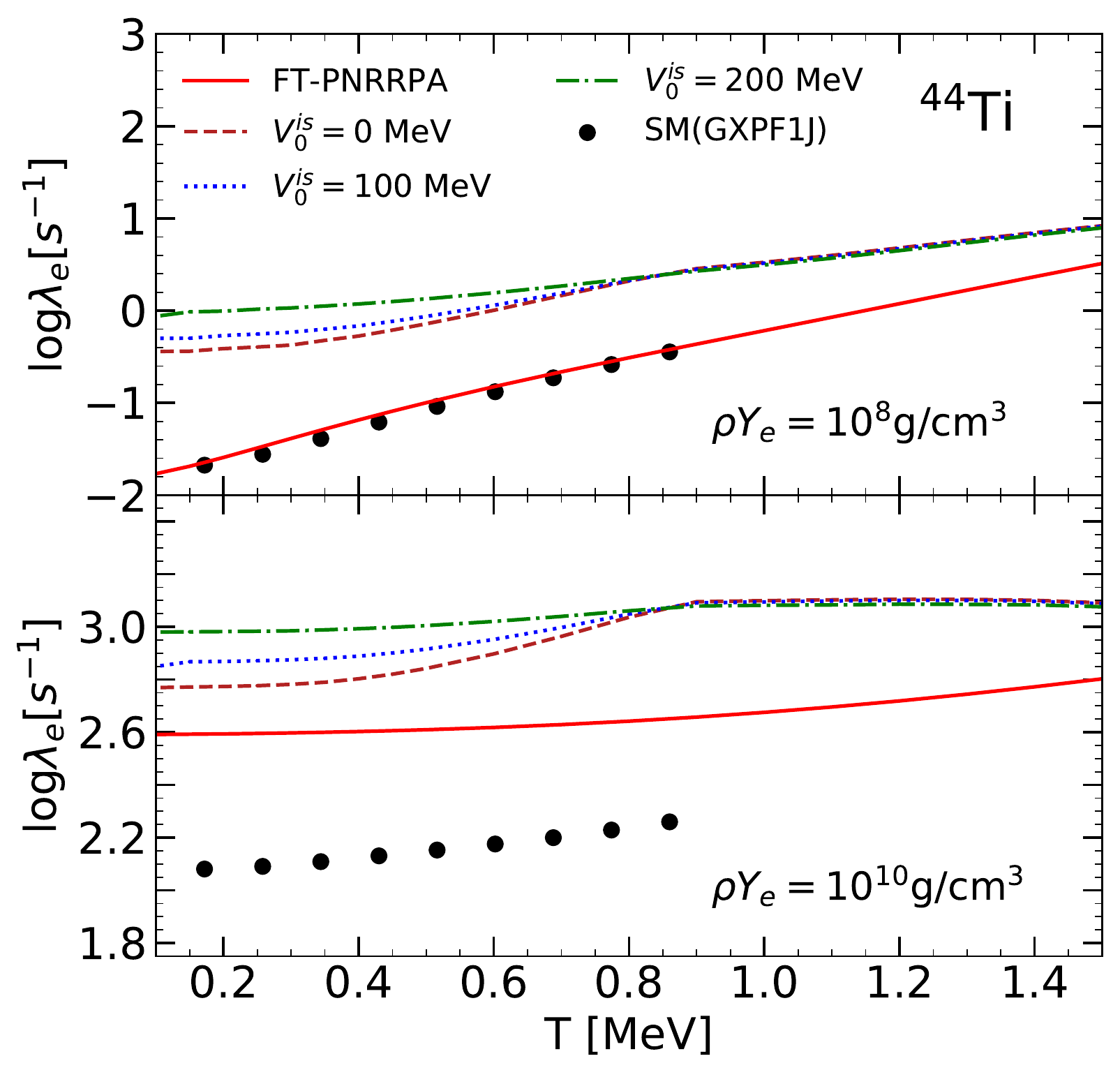}
\caption{Electron capture rates $\lambda_e$ for ${}^{44}$Ti as a function of temperature T for densities $\rho Y_e = 10^8$ and $10^{10}$ g/cm${}^3$. The FT-PNRRPA results (red solid line) are shown in comparison with the FT-PNRQRPA calculations using different values for the isoscalar pairing strength $V_0^{is}$. Results of shell-model (SM) calculations with GXPF1J interaction \cite{PhysRevC.83.044619,PhysRevC.65.061301, Mori_2016,SUZUKI2019} are also shown (black dots).}\label{fig:rates_ti44.png}
  \end{center}
\end{figure}

In Fig. \ref{fig:rates_ti44.png}, we display the results for the electron capture rates for ${}^{44}$Ti with increasing temperature.
The calculations are performed using both the FT-PNRRPA (red full line) and FT-PNRQRPA with $V_0^{is} = 0,100$ and 200 MeV to study the sensitivity of the results to the isoscalar pairing strength. 
We select the temperature interval as T=0-1.5 MeV and densities $\rho Y_e = 10^8$ and $10^{10}$ g/cm${}^3$, which are relevant for the evolution of core-collapse supernovae \cite{JUODAGALVIS2010454, Sullivan_2015, PhysRevC.86.015809}. We find that the calculations using the FT-PNRQRPA predict higher EC rates compared to the FT-PNRRPA results for $\rho Y_e = 10^8$ and $\rho Y_e = 10^{10}$ g/cm${}^3$. Below the critical temperature for protons, larger isoscalar pairing strength value $V_0^{is}$ produces larger rates. This result is also consistent with previous discussions on the GT${}^+$ strength and EC cross sections.
For $\rho Y_e = 10^{8}$ g/cm${}^3$, both the FT-PNRRPA and FT-PNRQRPA predict increasing EC rates with increasing temperature. Also, the difference between the two models decreases due to the vanishing of pairing properties with increasing temperature. In Fig. \ref{fig:rates_ti44.png}, we also present the shell-model (SM) results using the GXPF1J interaction (black full circles) for comparison \cite{PhysRevC.83.044619,PhysRevC.65.061301,Mori_2016,SUZUKI2019}. The FT-PNRRPA results are in good agreement with the shell-model calculations. Although the EC rates are overestimated using the FT-PNRQRPA compared to the shell model, it is seen that the behavior of the EC rates is compatible using both models. For higher electron densities, $\rho Y_e = 10^{10}$ g/cm${}^3$, rates are almost independent of temperature. By increasing temperature, the FT-PNRRPA calculation predicts slowly increasing EC rates. 
Using the FT-PNRQRPA, the behavior of the EC rates depends on the isoscalar pairing strength below the critical temperature for protons. We obtain a steep increase in the EC rates for the calculations without the isoscalar pairing, whereas it gradually increases for larger values of the isoscalar pairing strength. For T$>$0.9 MeV, the EC rates start to decrease slowly with increasing temperature.

The results obtained in Fig. \ref{fig:rates_ti44.png} can be explained by using the Eq. (\ref{eq:kinematics}) given in Section \ref{sec:formalism}. According to the Eq. (\ref{eq:kinematics}), we have a kinematic constraint on $E_{(Q)RPA}$ excitation energy (neglecting the neutrino mass)
\begin{equation}
E_e - E_{(Q)RPA} - \Delta_{np} (- \lambda_n + \lambda_p) > 0,
\label{100}
\end{equation}
where parentheses indicate that for the QRPA calculations additional subtraction of the neutron-proton chemical potential difference $\lambda_n - \lambda_p$ is needed. 
For electron energy $E_e$, we can use chemical potential of the electron $\lambda_e$, and by rearranging Eq. (\ref{100}) one obtains
\begin{equation}\label{eq:kin_cons}
E_{(Q)RPA} < \lambda_e -  \Delta_{np} (- \lambda_n + \lambda_p).
\end{equation}
According to Eq. (\ref{100}), we have a condition on the FT-PNR(Q)RPA excitation energies stemming from kinematics.
To compare the FT-PNRRPA excitation energies with the FT-PNRQRPA ones, we have to add the difference between neutron and proton chemical potentials ($\lambda_n - \lambda_p$) to the FT-PNRQRPA excitation energies and the above condition reduces to
\begin{equation}\label{eq:new_kin_cons}
E_{RPA} < \lambda_e -  \Delta_{np}.
\end{equation}

\begin{figure}[h!]
\includegraphics[width=1\linewidth,clip=true]{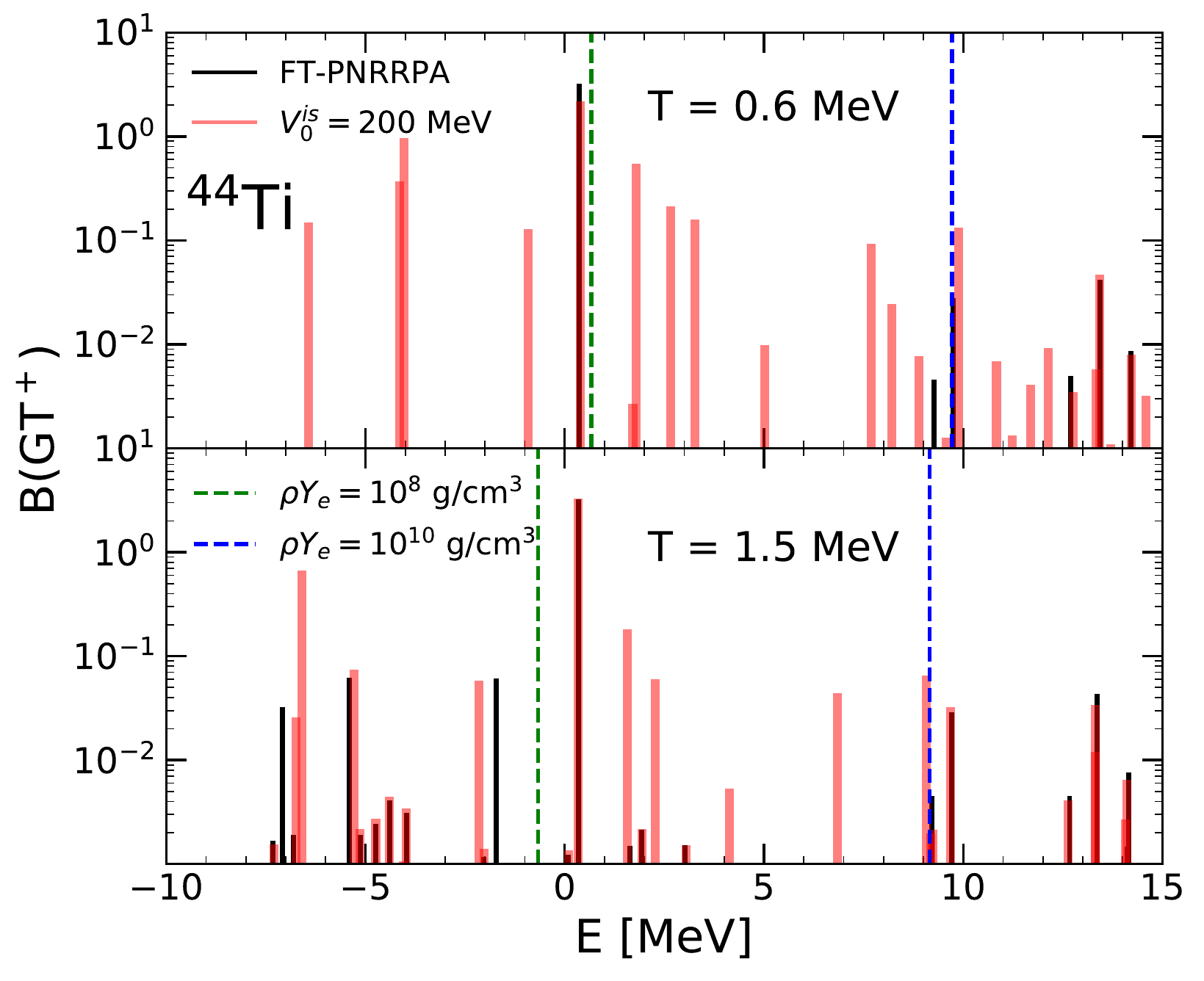}
\caption{Upper panel: The GT${}^+$ transition strength distributions for ${}^{44}$Ti calculated using the FT-PNRQRPA with $V_0^{is}$=200 MeV (red solid line) at T=0.6 MeV. The FT-PNRRPA results (black solid line) are also shown for comparison. Lower panel: The same but for T=1.5 MeV. Green dashed line is the upper limit on the excitation energy according to Eq. (\ref{eq:kin_cons}) for the electron chemical potential at $\rho Y_e = 10^8$ g/cm${}^3$ and blue dashed line is the same limit for the electron chemical potential at $\rho Y_e = 10^{10}$ g/cm${}^3$. See the text for details.}\label{fig:gtr_ti44_t06}
\end{figure}

In Fig. \ref{fig:gtr_ti44_t06}, the GT${}^+$  strength for ${}^{44}$Ti is displayed at T=0.6 and 1.5 MeV. The calculations are performed using the FT-PNRRPA and FT-PNRQRPA by fixing the isoscalar pairing strength to $V_0^{is}=200$ MeV. As can be seen from the upper panel of Fig. \ref{fig:gtr_ti44_t06}, the inclusion of pairing correlations unblocks the GT${}^+$ transitions and strength becomes considerably fragmented with the FT-PNRQRPA. Consequently, we obtain higher cross sections and higher EC rates. The GT${}^+$ excitation strength for ${}^{44}$Ti is also shown in the lower panel of Fig. \ref{fig:gtr_ti44_t06} at T=1.5 MeV. The results are displayed for the FT-PNRQRPA and FT-PNRRPA. Although the pairing effects are quite weak for ${}^{44}$Ti at T=1.5 MeV, we still have small differences in the predictions of the FT-PNRRPA and FT-PNRQRPA. The limiting values for the EC process (see Eq. (\ref{eq:new_kin_cons})), are also displayed on top of the GT${}^+$ excitation strength for $\rho Y_e = 10^8$ g/cm${}^3$ ($E_{QRPA} < 0.67$ MeV) and $\rho Y_e = 10^{10}$ g/cm${}^3$ ($E_{QRPA} < 9.72$ MeV). At higher value of chemical potential (obtained at $\rho Y_e = 10^{10}$ g/cm${}^3$), incoming electrons can excite more GT${}^+$ transitions, hence we obtain larger rates at higher stellar densities ($\rho Y_e$). Since the limit values for the EC process are quite low for $\rho Y_e = 10^8$ g/cm${}^3$, the EC rates are
more sensitive to the the structure of the low-energy states. By increasing temperature, 
the number of excited states increases considerably, resulting in gradual
increase of the the EC rates.
A similar explanation follows for $\rho Y_e = 10^{10}$ g/cm${}^3$, where most of the GT${}^+$ strength is excited due to the large electron chemical potential, and rates become almost independent on the increase of temperature.

For temperatures above $0.9$ MeV the rates become also less dependent on the pairing effects, which confirms the trends found in Fig. \ref{fig:gtr_ti44}, where for temperatures above the proton pairing collapse, the GT${}^+$ peaks for different values of $V_0^{is}$ almost match. The EC rates calculated using the FT-PNRRPA and FT-PNRQRPA (for $V_0^{is}=200$ MeV) 
differ by more than one order of magnitude for temperature below proton critical temperature
for $\rho Y_e = 10^8$ g/cm${}^3$ and by factor $\sim 2.5$ for $\rho Y_e = 10^{10}$ g/cm${}^3$. 
We note that in comparison to the shell-model, the FT-PNRQRPA
provides a self-consistent approach that allows a systematic description of the EC rates for
all nuclei of interest for supernova simulations, including complete description of all relevant multipole transitions.

\subsection{\texorpdfstring{$^{56}$Fe} {} nucleus}

As the next case we study $^{56}$Fe, the main ingredient in the core of the massive star 
at the end of hydrostatic burning. The electron capture process on iron group nuclei
plays an important role to set the conditions for the core collapse~\cite{RevModPhys.75.819}.  
Therefore, it is interesting to explore
the behavior of the GT${}^+$ strength as well as the EC cross sections and rates using the FT-PNRQRPA.
For the ground-state calculations, the monopole pairing constant is determined as $G_{n(p)}=29.6(30.6)$ MeV/A. It is found that the pairing collapse occurs at T$_c$ = 0.98 (0.85) MeV for neutron (proton) states. Therefore, our calculations do not have contributions from the pairing correlations for T $>$ 1 MeV.

\begin{figure}[!ht]
  \begin{center}
\includegraphics[width=1\linewidth,clip=true]{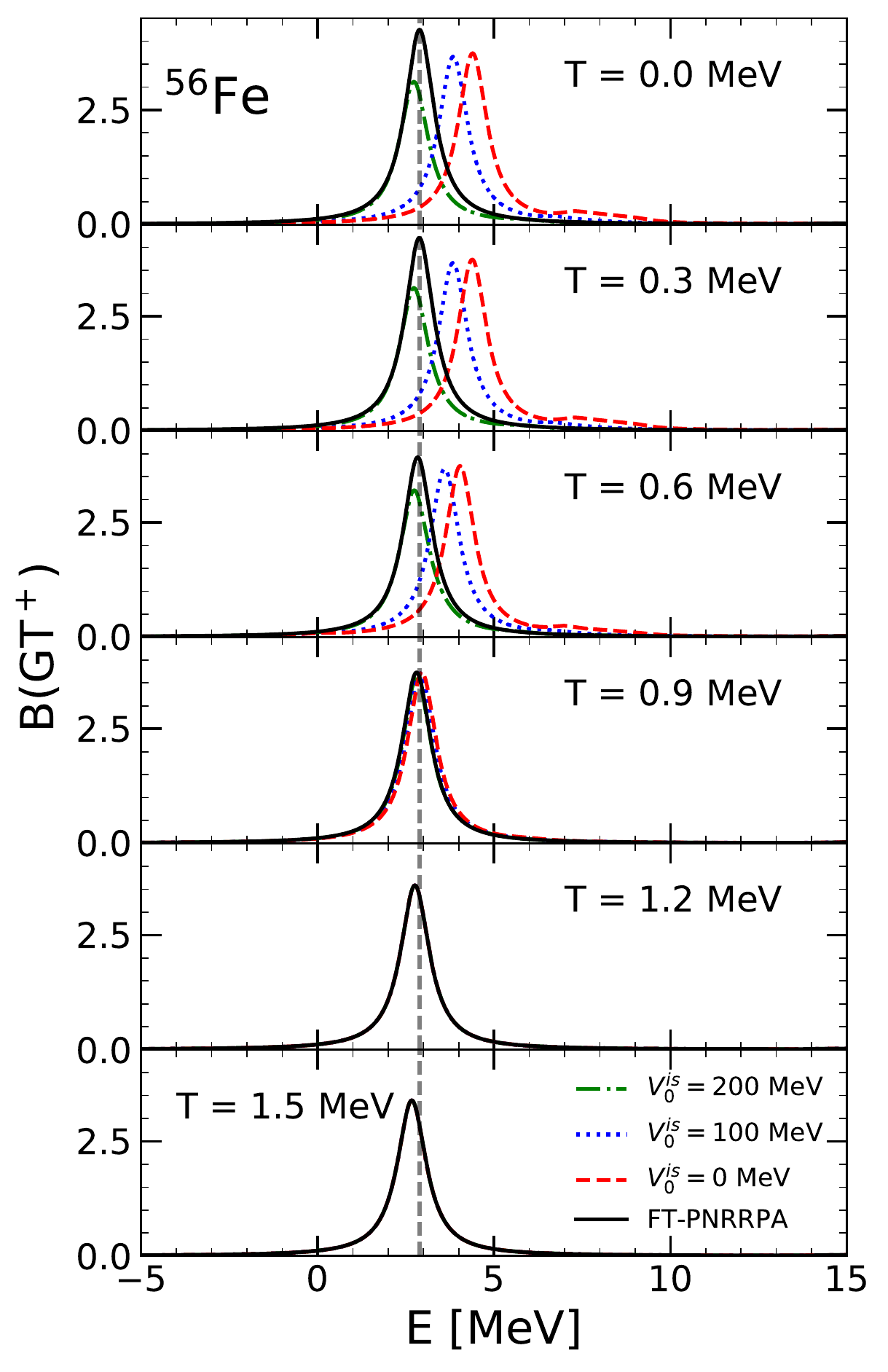}
\caption{Same as Fig. \ref{fig:gtr_ti44} but for ${}^{56}$Fe.}\label{fig:gtr_56fe}
  \end{center}
\end{figure}

Figure \ref{fig:gtr_56fe} shows the GT${}^+$ transition strength for ${}^{56}$Fe. The GT${}^+$ strength is calculated using the FT-PNRQRPA for different values of isoscalar pairing strength at T = 0, 0.3, 0.6, 0.9, 1.2 and 1.5 MeV. The FT-PNRRPA results are also presented for comparison.
At T = 0 MeV, the main FT-PNRRPA peak is found at E = 3.30 MeV  with B(GT${}^+$) = 6.68. The main contribution to its strength comes from $(\pi 1f_{7/2}, \nu 1f_{5/2})$ transition.

In the case of the FT-PNRQRPA calculations at T=0 limit, by increasing the isoscalar pairing strength $V_0^{is}$, the GT${}^+$ peak shifts to lower excitation energies due to the attractive nature of the isoscalar pairing. For the largest value of the isoscalar pairing strength ($V_0^{is} = 200$ MeV), the GT${}^+$ peak is found at $E = 2.73$ MeV with B(GT${}^+$) = 4.89. The main contribution of this state comes from $(\pi 1f_{7/2}, \nu 1f_{5/2})$ configuration, while some strength also comes from $(\pi 1f_{5/2}, \nu 1f_{7/2})$ transition. However, the strength of the peak slightly decreases due to the incoherent contribution of these transitions.

The evolution of the main peak with increasing temperature is shown in Tab. \ref{tab:strengths_56fe} using the FT-PNRRPA and FT-PNRQRPA. Using the FT-PNRRPA, the energy and strength of the main peak decreases gradually. However, this behavior changes with the inclusion of the pairing correlations.
Using the FT-PNRQRPA, the GT${}^+$ peak shifts slightly upward and
its strength increases from 4.89 to 5.69 up to T=0.9 MeV. Although the pairing effects weaken with increasing temperature, the modifications of the main peak reduce compared to the FT-PNRRPA due to the interplay between pairing and temperature effects. While the excited states are pushed down at finite temperatures due to the decrease in the two q.p. energies and residual particle-hole interaction, the reducing impact of the attractive residual isoscalar pairing also slows down the shift of the excited states to the lower energies. Therefore, the excited states are more stable against the changes in the temperature, at least up to the critical temperatures, compared to the FT-PNRRPA (see Ref. \cite{yksel2019gamowteller}).
By further increasing the temperature, pairing correlations disappear and the FT-PNRQRPA and FT-PNRRPA results completely agree as can be seen from Fig. \ref{fig:gtr_56fe} and Tab. \ref{tab:strengths_56fe}.

\begin{table}
\caption{The excitation energy and strength B(GT${}^+$) of the main peaks for ${}^{56}$Fe. The calculations are performed using the FT-PNRQRPA ($V_0^{is} = 200$ MeV) and FT-PNRRPA with increasing temperature. }\label{tab:strengths_56fe}
\begin{center}
\begin{tabular}{@{\extracolsep{4pt}}ccccccccc@{}}\hline\hline \\ [-1.5ex]
 &DD-ME2& \multicolumn{2}{c}{FT-PNRQRPA} & &\multicolumn{2}{c}{FT-PNRRPA}  \\ \cline{3-4} \cline{5-7} \\[-1.ex] 
 & T [MeV]& E [MeV]& B(GT${}^+$) &  & E [MeV]& B(GT${}^+$) \\ [1.ex] 
\hline
 \\ [-1ex]
&0.0 & 2.73 & 4.89 && 2.89 & 6.68 \\
&0.3 & 2.73 & 4.89 && 2.88 & 6.67 \\
&0.6 & 2.73 & 4.92 && 2.84 & 6.17 \\
&0.9 & 2.81 & 5.69 && 2.80 & 5.90 \\
&1.2 & 2.76 & 5.63 && 2.76 & 5.63 \\
&1.5 & 2.67 & 5.33 && 2.67 & 5.33 \\
\hline\hline
\end{tabular}
\end{center}
\vspace{-7mm}
\end{table}

\begin{figure*}[!ht]
  \begin{center}
\includegraphics[width=1\linewidth,clip=true]{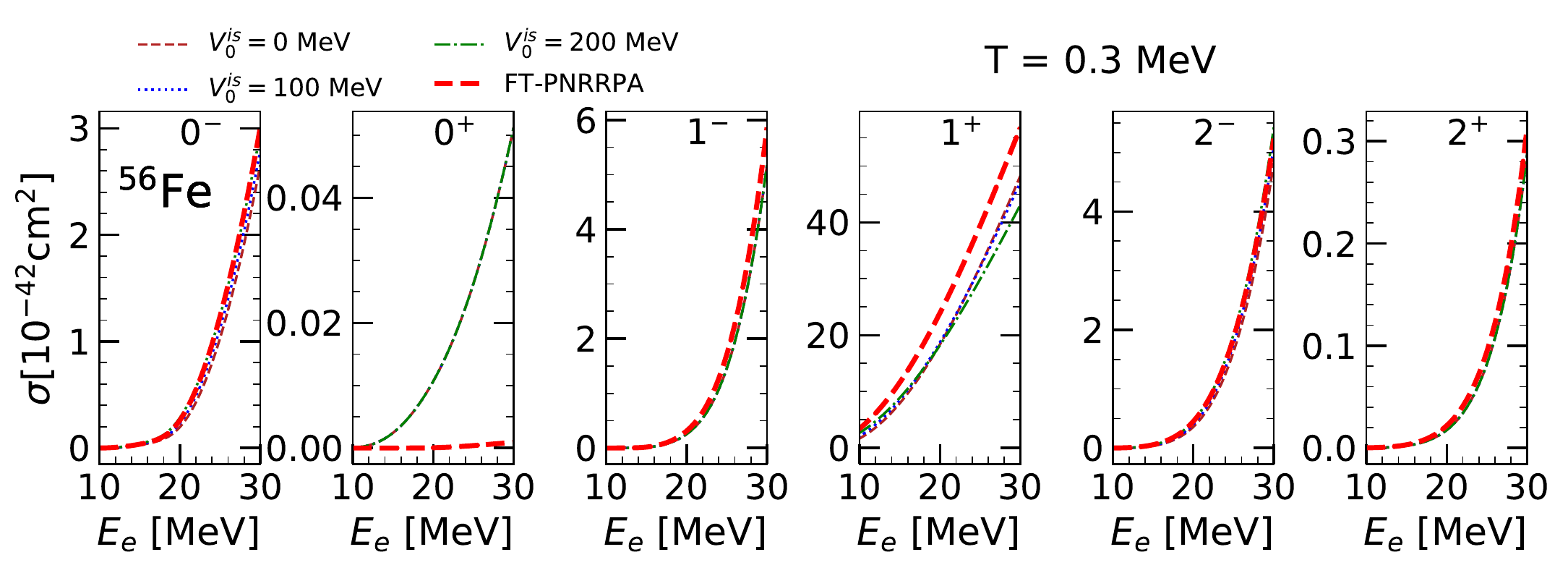}
\caption{Same as in Fig. \ref{fig:cross_sec_multipoles_ti44} but for ${}^{56}$Fe at T=0.3 MeV. }\label{fig:cross_sec_multipoles_fe56}
  \end{center}
\end{figure*}

Previous studies showed that the $1^+$ multipole gives the largest contribution to the total EC cross section for ${}^{56}$Fe \cite{PhysRevC.100.025801, PhysRevC.83.045807}. 
To verify this, we decompose the total electron capture cross section to multipoles up to $J=2$ for both positive and negative parities. 
The results are displayed in Fig. \ref{fig:cross_sec_multipoles_fe56} using the FT-PNRRPA and FT-PNRQRPA at T = 0.3 MeV. Clearly, the largest contribution to the EC cross section comes from $1^+$ multipole for both models. At low energies of incident electron the $1^+$ multipole operator reduces to GT${}^+$ operator \cite{PhysRevC.100.025801}. Therefore, the structure of the $1^+$ cross section is closely related to the GT${}^+$ strength distribution.
In Fig. \ref{fig:gtr_56fe}, one can observe that for temperatures below the pairing collapse (T $<$ 1 MeV), the FT-PNRRPA predicts more strength for the GT${}^+$ peak. This has a direct impact on the EC cross section results, and for $1^+$ multipole the FT-PNRRPA predicts larger values for all incident electron energies. Using the FT-PNRQRPA, the EC cross section for the $1^+$ multipole does not display sensitivity to the changes in the isoscalar pairing strength at lower electron energies. However, for ($E_e \sim 30$ MeV) the strongest GT${}^+$ transition is also excited by the electrons. In this case, the behavior of the EC cross section at a given temperature is mainly related to the main GT${}^+$ peak properties. Following the behavior of the GT${}^+$ peak shown in  Fig. \ref{fig:gtr_56fe}, with the increasing strength of the isoscalar pairing, the EC cross section for the $1^+$ multipole becomes reduced.

In addition to $1^+$ multipole, spin-dipole transitions $0^-, 1^-$ and $2^-$ also have non-negligible impact on the EC cross section. However, both the FT-PNRRPA and the FT-PNRQRPA predict similar results for those multipoles, with only small deviations. It is seen that $2^+$ and $0^+$ have the smallest contribution to the cross section. Similar to the findings in ${}^{44}$Ti, the FT-PNRQRPA predicts much larger contribution from $0^+$ multipole (Fermi transition) compared to the FT-PNRRPA. Since we have no significant temperature effects at T=0.3 MeV, the difference in the cross sections occurs due to the unblocking effect of the pairing correlations.

\begin{figure}
  \begin{center}
\includegraphics[width=1\linewidth,clip=true]{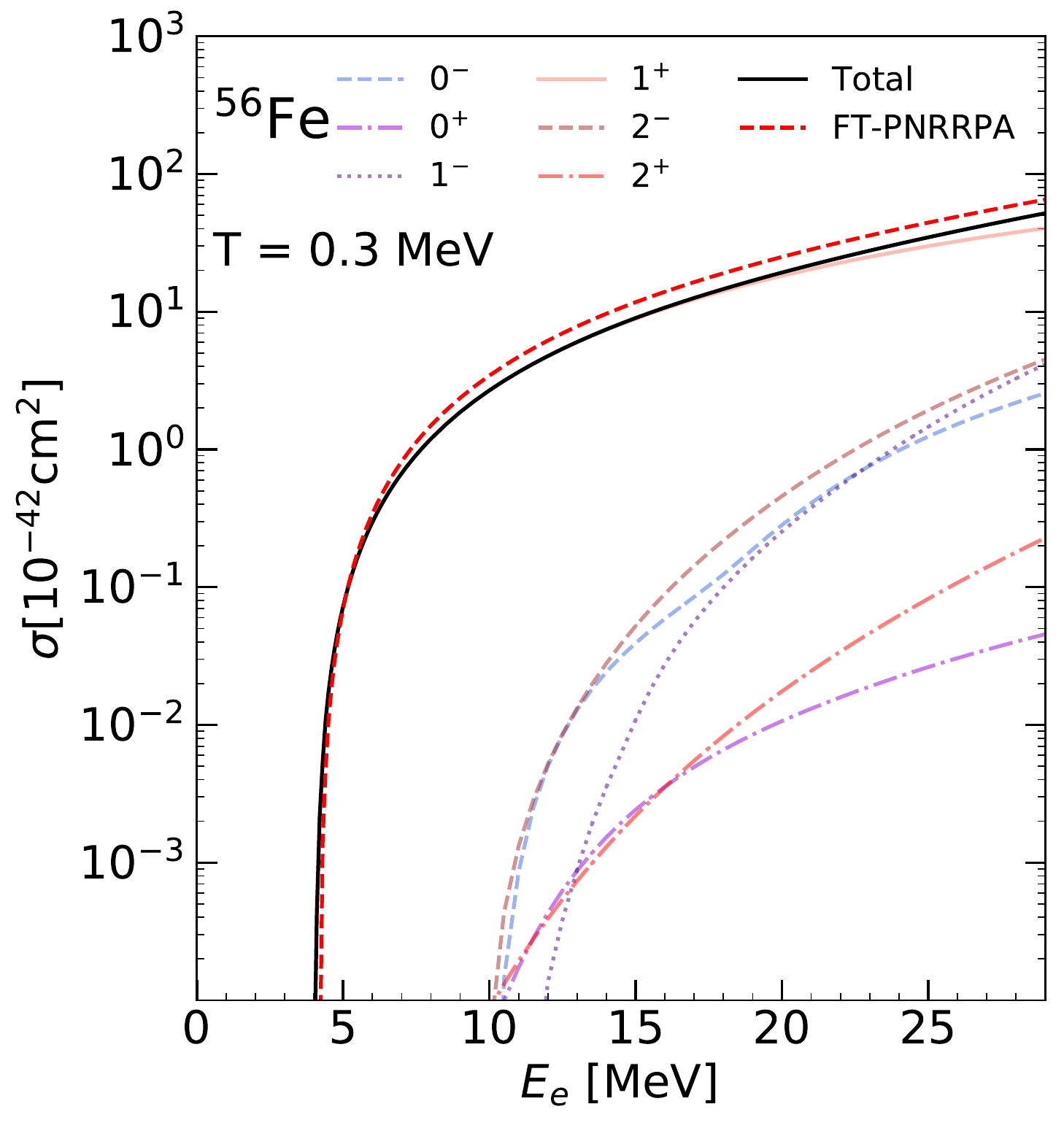}
\caption{Same as in Fig. \ref{fig:cross_sec_multipoles_ti44} but for ${}^{56}$Fe at T=0.3 MeV. }\label{fig:cross_section_T3FE56.png}
  \end{center}
\end{figure}

In Fig. \ref{fig:cross_section_T3FE56.png}, the EC cross section results are displayed for all multipoles at T=0.3 MeV. The isoscalar pairing strength is taken as $V_0^{is} = 200$ MeV for the FT-PNRQRPA calculations. At higher energies of incident electron, the FT-PNRRPA predicts larger total cross section, as it was concluded from our discussion on $1^+$ multipole results. It can be clearly seen that $1^+$ dominates the EC cross sections up to high electron energies ($E_e \sim 25$ MeV) where contributions from the forbidden multipoles ($0^-$, $1^-$, $2^-$) become non-negligible.

\begin{figure}
  \begin{center}
\includegraphics[width=1\linewidth,clip=true]{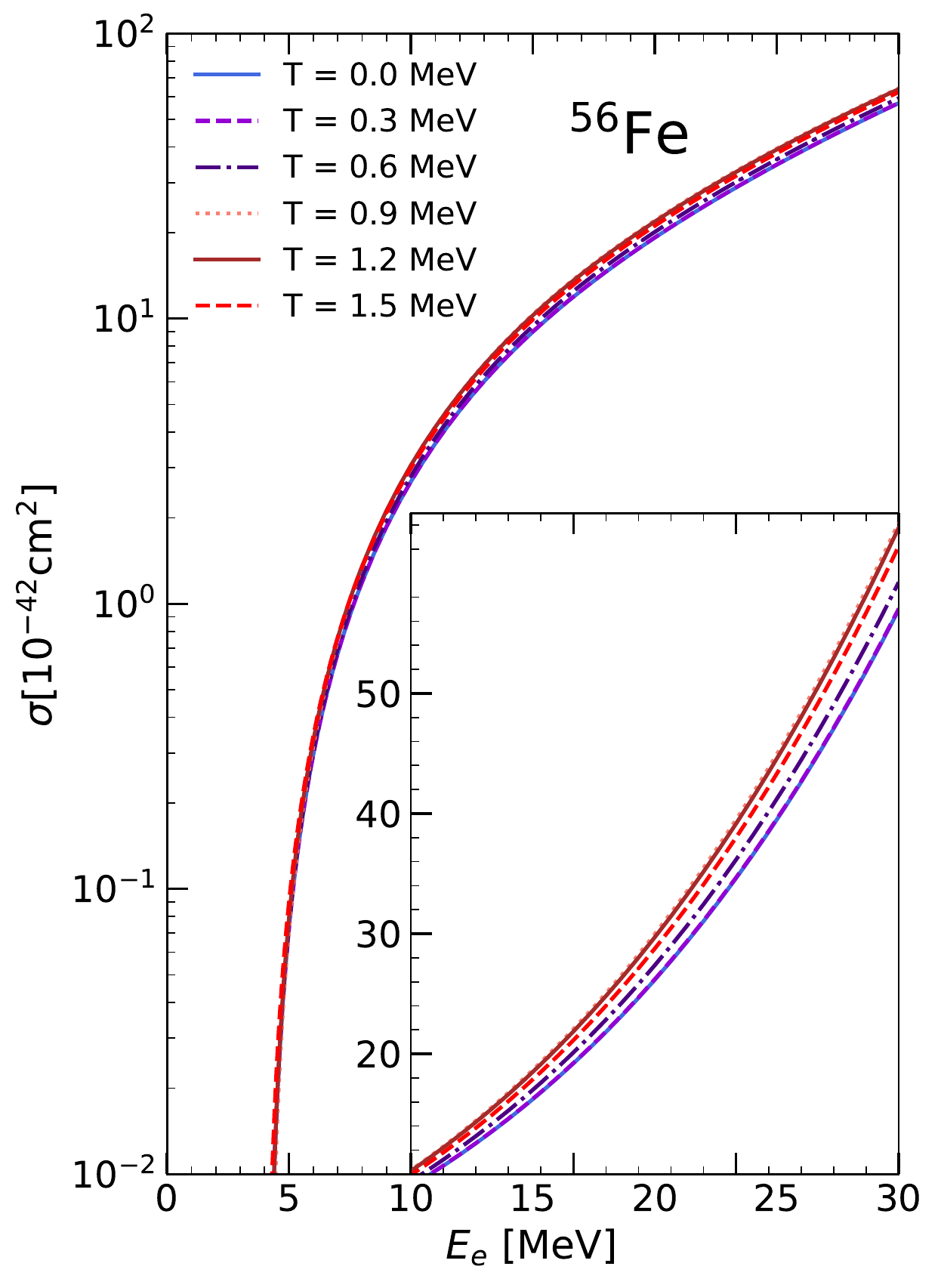}
\caption{Same as in Fig. \ref{fig:temp_ti44.png} but for ${}^{56}$Fe. }\label{fig:temp_fe56.png}
  \end{center}
\end{figure}

In Fig. \ref{fig:temp_fe56.png}, we show the dependence of EC cross section on temperature using the FT-PNRQRPA with $V_0^{is} = 200$ MeV. In the inner panel of the figure, the range of the incident electron energies are limited to $E_e = 10-30$ MeV to better visualize dependence on the temperature. Up to T=0.3 MeV, the cross section displays no change. 
By increasing the temperature up to T=0.9 MeV, the cross sections increase due to the increase in the strength of the GT${}^+$ peak (see Tab. \ref{tab:strengths_56fe}). 
Above the critical temperatures, the pairing effects disappear, whereas no significant temperature unblocking occurs for the considered temperature range.
Therefore, the total GT${}^+$ strength and cross sections decrease with increasing temperature. 
Calculations using the FT-PNRRPA without the pairing correlations predict a gradual decrease in the excitation energies and strength (see Tab. \ref{tab:strengths_56fe}). Therefore, the EC cross sections decrease below T=0.9 MeV as compared to the calculations using the FT-PNRQRPA. This is a proof of the importance of including pairing correlations in the EC calculations below the critical temperature.

\begin{figure}
  \begin{center}
\includegraphics[width=1\linewidth,clip=true]{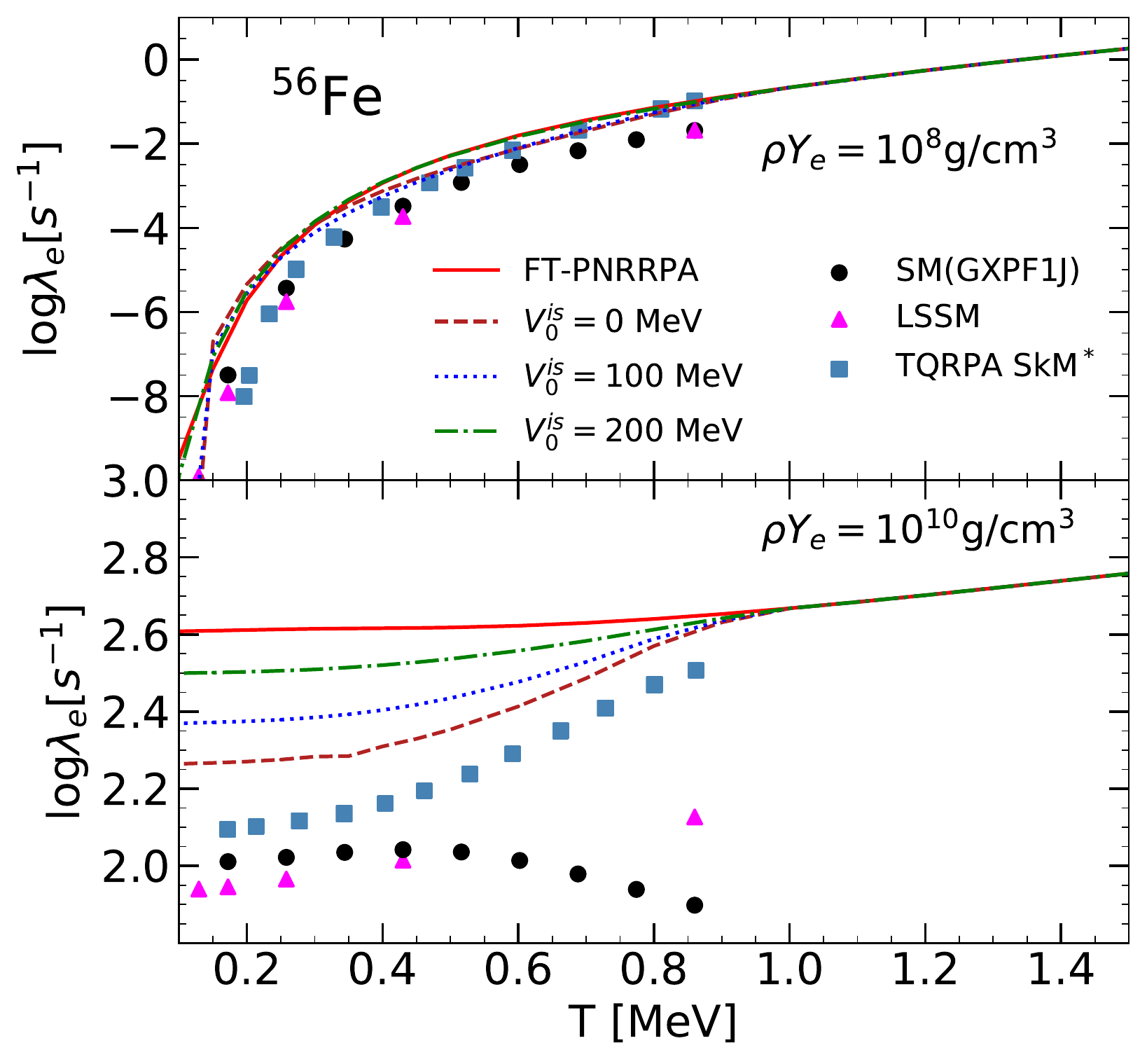}
\caption{Electron capture rates $\lambda_e$ for ${}^{56}$Fe with respect to temperature T for densities $\rho Y_e = 10^8$ and $10^{10}$ g/cm${}^3$. The results for the FT-PNRRPA (red solid line) are shown together with the FT-PNRQRPA calculations for different strengths of isoscalar pairing $V_0^{is}$. The EC rates using the TQRPA (blue squares) from Ref. \cite{PhysRevC.100.025801}, the shell-model (SM) with GXPF1J interaction (black dots)\cite{PhysRevC.83.044619,PhysRevC.65.061301,Mori_2016,SUZUKI2019}
and LSSM calculations (purple triangles) \cite{LANGANKE2000481} are also shown for comparison.} 
\label{fig:rates_fe56.png}
 \end{center}
\end{figure}

Results for the EC rates for ${}^{56}$Fe are shown in Fig. \ref{fig:rates_fe56.png} for two cases of densities $\rho Y_e = 10^8$ and $10^{10}$ g/cm${}^3$. The FT-PNRQRPA calculations using $V_0^{is} = 0,100$ and 200 MeV are shown together with the FT-PNRRPA results (red solid line). For comparison, the EC rates from other model calculations are also displayed: thermal QRPA (TQRPA) with the Skyrme SkM${}^*$ interaction  (blue squares) \cite{PhysRevC.100.025801}, the shell-model (SM) with GXPF1J interaction (black dots) \cite{PhysRevC.83.044619,PhysRevC.65.061301, Mori_2016,SUZUKI2019} and the LSSM calculations (purple triangles) \cite{LANGANKE2000481}. 
At $\rho Y_e = 10^8$ g/cm${}^3$, the EC rates are almost independent of the isoscalar pairing strength $V_0^{is}$ and increase with increasing temperature. Both the FT-PNRQRPA and FT-PNRRPA predict similar results and agree with the TQRPA calculations. Our calculations also show very good agreement with the LSSM calculations~\cite{LANGANKE2000481} and the shell-model calculations with GXPF1J interaction ~\cite{PhysRevC.83.044619,PhysRevC.65.061301, Mori_2016,SUZUKI2019} in the whole temperature range. Since we have low electron chemical potential ($\lambda_e \sim 2$ MeV), the main GT$^+$ peak from Fig. \ref{fig:gtr_56fe} is not included in the EC rates calculation. Therefore, small EC rates are obtained at lower temperatures for $\rho Y_e = 10^8$ g/cm${}^3$.  By increasing the temperature (i) the main peak shifts to lower excitation energies and (ii) additional peaks appear at low (even negative) excitation energies. Consequently, the EC rates start to increase with increasing temperature.

 At $\rho Y_e = 10^{10}$ g/cm${}^3$ the electron chemical potential is large enough ($\lambda_e \sim 10$ MeV) to excite most of the GT${}^+$ strength. Therefore, we obtain high EC rates compared to the previous case of lower density. The FT-PNRRPA predicts larger EC rates than the FT-PNRQRPA ones, irrespective of the isoscalar pairing strength $V_0^{is}$. As indicated in our analysis of the GT${}^+$ strength in Fig. \ref{fig:gtr_56fe}, for the higher value of the isoscalar pairing strength $V_0^{is}$, the main peak is pushed to lower excitation energies, and GT${}^+$ strength is more easily excited by incoming electrons. 
When compared with the TQRPA calculations, the FT-PNRQRPA rates are
somewhat larger. However, one should note that the TQRPA is rather different approach than the FT-PNRQRPA. The TQRPA calculations are based on the non-relativistic Skyrme functional (SkM${}^*$), without the isoscalar pairing in the residual interaction of the TQRPA. In our case, the relativistic functional (DD-ME2) is employed in calculations, supplemented with the $T=0$ pairing in the residual interaction. Our results are in reasonable agreement with the LSSM rates~\cite{LANGANKE2000481}, while the shell model calculations based on GXPF1J interaction ~\cite{Mori_2016,SUZUKI2019} show different trend with increasing temperature.
We note that the SM rates ~\cite{Mori_2016,SUZUKI2019} start to decrease above
T$\sim$0.5 MeV because these calculations also include additional transitions from other excited states except 0$^+$ ground state and first 2$^+$ states~\cite{SUZUKI2019}.
The EC rates calculated with the FT-PNRQRPA  display weak dependence on temperature for high stellar densities.

We should mention that only two-quasiparticle excitations are considered within the FT-PNRQRPA calculations, while the shell-model calculations take into account complex configurations, thus better predicting the fragmentation in the excitation strength. Although at high stellar densities the EC rates slowly vary at higher temperature due to large chemical potential, and depend on total GT${}^+$ strength, this weak temperature dependence is sensitive to the fragmented strength. To confirm this we can explore the EC rates calculated with different values of isoscalar pairing in Fig. \ref{fig:rates_fe56.png}. Even though the chemical potential is high enough to excite all of the GT${}^+$ strength, the variation of isoscalar pairing strength produces small variations in the temperature dependence of the rates. Similar conclusion applies to the shell-model strength compared to the FT-PNRQRPA.

\section{Conclusion}\label{sec:conclusion}

In this work, we have studied the electron capture cross sections and rates in stellar environment, based on the relativistic energy density functional to describe relevant nuclear properties and transitions. In comparison to previous studies based on energy density functionals, we introduced a framework to describe the EC process for the first time by including both the finite temperature and nuclear pairing effects. The FT-HBCS model was employed to calculate nuclear ground-state properties and the FT-PNRQRPA\cite{PhysRevC.96.024303} was 
used to describe the relevant nuclear excitations in the charge-exchange channel.
Our model is self-consistent in a sense that the same relativistic energy density functional (DD-ME2) \cite{PhysRevC.71.024312} is used in the ground-state and excited-state calculations. The pairing interaction is included both in the ground-state calculations and in the residual FT-PNRQRPA interaction. In the latter case, the isoscalar pairing interaction is also included, that necessitates further constraint by the experimental data. Rather than constraining its value, in this work we have explored the sensitivity of the results on the $T=0$ pairing interaction strength parameter.

We first analyzed the GT${}^+$ strength distributions in ${}^{44}$Ti and ${}^{56}$Fe at finite temperatures. Then, the EC cross sections and rates are investigated, by including transitions up to $J^\pi = 0^\pm, 1^\pm, 2^\pm$ multipoles. We mainly focused on the role of the pairing and temperature effects on the GT${}^+$ transitions and EC calculations. It is shown that the isoscalar pairing plays an important role below the critical temperature and together with the temperature effects leads to the unblocking of quasiparticle states and new excitation channels become possible.

By using a range of values for the isoscalar pairing strength $V_0^{is}$, different results for the EC cross sections and rates are obtained, however, all results are within the same order of the magnitude. Nevertheless, a better estimation of the isoscalar pairing strength constrained by other experimental data is needed for future study. 
By increasing temperature, additional peaks in the GT${}^+$ strength appear and excited states shift downward. Compared to the results without the pairing correlations, the downward shift of the excited states is mainly slowed down with the inclusion of pairing below the critical temperatures where the pairing correlations vanish. As discussed in Ref.~\cite{yksel2019gamowteller}, this behavior is a result of the interplay between the pairing and temperature effects below the critical temperatures.
It is also shown that the inclusion of the pairing correlations along with the temperature can impact the EC cross sections and rates considerably, compared to the calculations without the pairing effects. 
The pairing correlations can alter the EC cross-sections up to the factor of two for 1$^+$ multipole in ${}^{44}$Ti. Model calculations also demonstrated the importance of including forbidden transitions in the description of EC process with increasing temperature. 
Our results are also in a qualitative agreement with the TQRPA \cite{PhysRevC.100.025801} and shell-model \cite{PhysRevC.83.044619,PhysRevC.65.061301, Mori_2016} calculations. Since temperatures below the pairing collapse are also important for the evolution of core-collapse supernovae, the theoretical framework introduced in this work represents a complete and consistent microscopic tool that can be readily applied to describe all EC rates relevant for the core-collapse supernovae simulations. 

The FT-PNRQRPA formalism in the charge-exchange channel can also be applied to beta-decay and neutrino-nucleus reactions at finite temperature, and hence provide a universal theoretical approach to weak-interaction processes important for the evolution of core-collapse supernovae. Since our model is based on the BCS theory, it cannot be
applied for exotic nuclei close to the drip lines where scattering to continuum becomes more important. Further improvements toward finite temperature Hartree-Bogoliubov theory \cite{PhysRevC.88.034308} that would successfully describe those nuclei, also including the nuclear deformation effects are currently under development.

\section{Acknowledgements}
We thank Toshio Suzuki for providing us the data from the shell model calculations and useful discussions.
This work is supported by the QuantiXLie Centre of Excellence, a project co-financed by the Croatian Government and European Union through the European Regional Development Fund, the Competitiveness and Cohesion Operational Programme (KK.01.1.1.01). This article is based upon work from the “ChETEC” COST Action (CA16117), supported by COST (European Cooperation in Science and Technology).
E. Y. acknowledges financial support from the Scientific
and Technological Research Council of Turkey (T\"{U}B\.{I}TAK) BIDEB-2219 Postdoctoral Research program.
Y. F. N. acknowledges the support from the Fundamental Research Funds for the Central Universities under Grant No. Lzujbky-2019-11. G.C. acknowledges funding from the European
Union's Horizon 2020 research and innovation program
under Grant No. 654002.

\bibliographystyle{apsrev4-1}
\bibliography{electron_capture_NP.bbl}

\begin{thebibliography}{60}%
\makeatletter
\providecommand \@ifxundefined [1]{%
 \@ifx{#1\undefined}
}%
\providecommand \@ifnum [1]{%
 \ifnum #1\expandafter \@firstoftwo
 \else \expandafter \@secondoftwo
 \fi
}%
\providecommand \@ifx [1]{%
 \ifx #1\expandafter \@firstoftwo
 \else \expandafter \@secondoftwo
 \fi
}%
\providecommand \natexlab [1]{#1}%
\providecommand \enquote  [1]{``#1''}%
\providecommand \bibnamefont  [1]{#1}%
\providecommand \bibfnamefont [1]{#1}%
\providecommand \citenamefont [1]{#1}%
\providecommand \href@noop [0]{\@secondoftwo}%
\providecommand \href [0]{\begingroup \@sanitize@url \@href}%
\providecommand \@href[1]{\@@startlink{#1}\@@href}%
\providecommand \@@href[1]{\endgroup#1\@@endlink}%
\providecommand \@sanitize@url [0]{\catcode `\\12\catcode `\$12\catcode
  `\&12\catcode `\#12\catcode `\^12\catcode `\_12\catcode `\%12\relax}%
\providecommand \@@startlink[1]{}%
\providecommand \@@endlink[0]{}%
\providecommand \url  [0]{\begingroup\@sanitize@url \@url }%
\providecommand \@url [1]{\endgroup\@href {#1}{\urlprefix }}%
\providecommand \urlprefix  [0]{URL }%
\providecommand \Eprint [0]{\href }%
\providecommand \doibase [0]{http://dx.doi.org/}%
\providecommand \selectlanguage [0]{\@gobble}%
\providecommand \bibinfo  [0]{\@secondoftwo}%
\providecommand \bibfield  [0]{\@secondoftwo}%
\providecommand \translation [1]{[#1]}%
\providecommand \BibitemOpen [0]{}%
\providecommand \bibitemStop [0]{}%
\providecommand \bibitemNoStop [0]{.\EOS\space}%
\providecommand \EOS [0]{\spacefactor3000\relax}%
\providecommand \BibitemShut  [1]{\csname bibitem#1\endcsname}%
\let\auto@bib@innerbib\@empty
\bibitem [{\citenamefont {Bethe}\ \emph {et~al.}(1979)\citenamefont {Bethe},
  \citenamefont {Brown}, \citenamefont {Applegate},\ and\ \citenamefont
  {Lattimer}}]{BETHE1979487}%
  \BibitemOpen
  \bibfield  {author} {\bibinfo {author} {\bibfnamefont {H.}~\bibnamefont
  {Bethe}}, \bibinfo {author} {\bibfnamefont {G.}~\bibnamefont {Brown}},
  \bibinfo {author} {\bibfnamefont {J.}~\bibnamefont {Applegate}}, \ and\
  \bibinfo {author} {\bibfnamefont {J.}~\bibnamefont {Lattimer}},\ }\href
  {\doibase https://doi.org/10.1016/0375-9474(79)90596-7} {\bibfield  {journal}
  {\bibinfo  {journal} {Nuclear Physics A}\ }\textbf {\bibinfo {volume}
  {324}},\ \bibinfo {pages} {487 } (\bibinfo {year} {1979})}\BibitemShut
  {NoStop}%
\bibitem [{\citenamefont {Janka}\ \emph {et~al.}(2007)\citenamefont {Janka},
  \citenamefont {Langanke}, \citenamefont {Marek}, \citenamefont
  {Mart{\'{\i}}nez-Pinedo},\ and\ \citenamefont {M{\"o}ller}}]{JANKA200738}%
  \BibitemOpen
  \bibfield  {author} {\bibinfo {author} {\bibfnamefont {H.-T.}\ \bibnamefont
  {Janka}}, \bibinfo {author} {\bibfnamefont {K.}~\bibnamefont {Langanke}},
  \bibinfo {author} {\bibfnamefont {A.}~\bibnamefont {Marek}}, \bibinfo
  {author} {\bibfnamefont {G.}~\bibnamefont {Mart{\'{\i}}nez-Pinedo}}, \ and\
  \bibinfo {author} {\bibfnamefont {B.}~\bibnamefont {M{\"o}ller}},\ }\href
  {\doibase https://doi.org/10.1016/j.physrep.2007.02.002} {\bibfield
  {journal} {\bibinfo  {journal} {Physics Reports}\ }\textbf {\bibinfo {volume}
  {442}},\ \bibinfo {pages} {38 } (\bibinfo {year} {2007})},\ \bibinfo {note}
  {the Hans Bethe Centennial Volume 1906-2006}\BibitemShut {NoStop}%
\bibitem [{\citenamefont {Bethe}\ and\ \citenamefont
  {Brown}(1985)}]{bethe_how}%
  \BibitemOpen
  \bibfield  {author} {\bibinfo {author} {\bibfnamefont {H.~A.}\ \bibnamefont
  {Bethe}}\ and\ \bibinfo {author} {\bibfnamefont {G.}~\bibnamefont {Brown}},\
  }\href {http://www.jstor.org/stable/24967637} {\bibfield  {journal} {\bibinfo
   {journal} {Scientific American}\ }\textbf {\bibinfo {volume} {252}},\
  \bibinfo {pages} {60} (\bibinfo {year} {1985})}\BibitemShut {NoStop}%
\bibitem [{\citenamefont {Fuller}\ \emph {et~al.}(1980)\citenamefont {Fuller},
  \citenamefont {Fowler},\ and\ \citenamefont {Newman}}]{fuller1980stellar}%
  \BibitemOpen
  \bibfield  {author} {\bibinfo {author} {\bibfnamefont {G.~M.}\ \bibnamefont
  {Fuller}}, \bibinfo {author} {\bibfnamefont {W.~A.}\ \bibnamefont {Fowler}},
  \ and\ \bibinfo {author} {\bibfnamefont {M.~J.}\ \bibnamefont {Newman}},\
  }\href {\doibase https://dx.doi.org/10.1086/190657} {\bibfield  {journal}
  {\bibinfo  {journal} {The Astrophysical Journal Supplement Series}\ }\textbf
  {\bibinfo {volume} {42}},\ \bibinfo {pages} {447} (\bibinfo {year}
  {1980})}\BibitemShut {NoStop}%
\bibitem [{\citenamefont {Fuller}\ \emph
  {et~al.}(1982{\natexlab{a}})\citenamefont {Fuller}, \citenamefont {Fowler},\
  and\ \citenamefont {Newman}}]{fuller1982stellar_part2}%
  \BibitemOpen
  \bibfield  {author} {\bibinfo {author} {\bibfnamefont {G.~M.}\ \bibnamefont
  {Fuller}}, \bibinfo {author} {\bibfnamefont {W.~A.}\ \bibnamefont {Fowler}},
  \ and\ \bibinfo {author} {\bibfnamefont {M.~J.}\ \bibnamefont {Newman}},\
  }\href {\doibase https://dx.doi.org/10.1086/190779} {\bibfield  {journal}
  {\bibinfo  {journal} {Astrophysical Journal Supplement Series}\ }\textbf
  {\bibinfo {volume} {48}},\ \bibinfo {pages} {279} (\bibinfo {year}
  {1982}{\natexlab{a}})}\BibitemShut {NoStop}%
\bibitem [{\citenamefont {Fuller}\ \emph
  {et~al.}(1982{\natexlab{b}})\citenamefont {Fuller}, \citenamefont {Fowler},\
  and\ \citenamefont {Newman}}]{fuller1982stellar_part3}%
  \BibitemOpen
  \bibfield  {author} {\bibinfo {author} {\bibfnamefont {G.~M.}\ \bibnamefont
  {Fuller}}, \bibinfo {author} {\bibfnamefont {W.}~\bibnamefont {Fowler}}, \
  and\ \bibinfo {author} {\bibfnamefont {M.}~\bibnamefont {Newman}},\ }\href
  {\doibase https://dx.doi.org/10.1086/159597} {\bibfield  {journal} {\bibinfo
  {journal} {Astrophysical Journal Supplement Series}\ }\textbf {\bibinfo
  {volume} {252}},\ \bibinfo {pages} {715} (\bibinfo {year}
  {1982}{\natexlab{b}})}\BibitemShut {NoStop}%
\bibitem [{\citenamefont {Fuller}\ \emph {et~al.}(1985)\citenamefont {Fuller},
  \citenamefont {Fowler},\ and\ \citenamefont {Newman}}]{fuller1985stellar}%
  \BibitemOpen
  \bibfield  {author} {\bibinfo {author} {\bibfnamefont {G.}~\bibnamefont
  {Fuller}}, \bibinfo {author} {\bibfnamefont {W.}~\bibnamefont {Fowler}}, \
  and\ \bibinfo {author} {\bibfnamefont {M.}~\bibnamefont {Newman}},\ }\href
  {\doibase https://dx.doi.org/10.1086/163208} {\bibfield  {journal} {\bibinfo
  {journal} {Astrophysical Journal Supplement Series}\ }\textbf {\bibinfo
  {volume} {293}},\ \bibinfo {pages} {1} (\bibinfo {year} {1985})}\BibitemShut
  {NoStop}%
\bibitem [{\citenamefont {Langanke}\ and\ \citenamefont
  {Mart{\'{\i}}nez-Pinedo}(2001)}]{LANGANKE20011}%
  \BibitemOpen
  \bibfield  {author} {\bibinfo {author} {\bibfnamefont {K.}~\bibnamefont
  {Langanke}}\ and\ \bibinfo {author} {\bibfnamefont {G.}~\bibnamefont
  {Mart{\'{\i}}nez-Pinedo}},\ }\href {\doibase
  https://doi.org/10.1006/adnd.2001.0865} {\bibfield  {journal} {\bibinfo
  {journal} {Atomic Data and Nuclear Data Tables}\ }\textbf {\bibinfo {volume}
  {79}},\ \bibinfo {pages} {1 } (\bibinfo {year} {2001})}\BibitemShut {NoStop}%
\bibitem [{\citenamefont {Honma}\ \emph {et~al.}(2002)\citenamefont {Honma},
  \citenamefont {Otsuka}, \citenamefont {Brown},\ and\ \citenamefont
  {Mizusaki}}]{PhysRevC.65.061301}%
  \BibitemOpen
  \bibfield  {author} {\bibinfo {author} {\bibfnamefont {M.}~\bibnamefont
  {Honma}}, \bibinfo {author} {\bibfnamefont {T.}~\bibnamefont {Otsuka}},
  \bibinfo {author} {\bibfnamefont {B.~A.}\ \bibnamefont {Brown}}, \ and\
  \bibinfo {author} {\bibfnamefont {T.}~\bibnamefont {Mizusaki}},\ }\href
  {\doibase 10.1103/PhysRevC.65.061301} {\bibfield  {journal} {\bibinfo
  {journal} {Phys. Rev. C}\ }\textbf {\bibinfo {volume} {65}},\ \bibinfo
  {pages} {061301(R)} (\bibinfo {year} {2002})}\BibitemShut {NoStop}%
\bibitem [{\citenamefont {Suzuki}\ \emph {et~al.}(2011)\citenamefont {Suzuki},
  \citenamefont {Honma}, \citenamefont {Mao}, \citenamefont {Otsuka},\ and\
  \citenamefont {Kajino}}]{PhysRevC.83.044619}%
  \BibitemOpen
  \bibfield  {author} {\bibinfo {author} {\bibfnamefont {T.}~\bibnamefont
  {Suzuki}}, \bibinfo {author} {\bibfnamefont {M.}~\bibnamefont {Honma}},
  \bibinfo {author} {\bibfnamefont {H.}~\bibnamefont {Mao}}, \bibinfo {author}
  {\bibfnamefont {T.}~\bibnamefont {Otsuka}}, \ and\ \bibinfo {author}
  {\bibfnamefont {T.}~\bibnamefont {Kajino}},\ }\href {\doibase
  10.1103/PhysRevC.83.044619} {\bibfield  {journal} {\bibinfo  {journal} {Phys.
  Rev. C}\ }\textbf {\bibinfo {volume} {83}},\ \bibinfo {pages} {044619}
  (\bibinfo {year} {2011})}\BibitemShut {NoStop}%
\bibitem [{\citenamefont {Mori}\ \emph {et~al.}(2016)\citenamefont {Mori},
  \citenamefont {Famiano}, \citenamefont {Kajino}, \citenamefont {Suzuki},
  \citenamefont {Hidaka}, \citenamefont {Honma}, \citenamefont {Iwamoto},
  \citenamefont {Nomoto},\ and\ \citenamefont {Otsuka}}]{Mori_2016}%
  \BibitemOpen
  \bibfield  {author} {\bibinfo {author} {\bibfnamefont {K.}~\bibnamefont
  {Mori}}, \bibinfo {author} {\bibfnamefont {M.~A.}\ \bibnamefont {Famiano}},
  \bibinfo {author} {\bibfnamefont {T.}~\bibnamefont {Kajino}}, \bibinfo
  {author} {\bibfnamefont {T.}~\bibnamefont {Suzuki}}, \bibinfo {author}
  {\bibfnamefont {J.}~\bibnamefont {Hidaka}}, \bibinfo {author} {\bibfnamefont
  {M.}~\bibnamefont {Honma}}, \bibinfo {author} {\bibfnamefont
  {K.}~\bibnamefont {Iwamoto}}, \bibinfo {author} {\bibfnamefont
  {K.}~\bibnamefont {Nomoto}}, \ and\ \bibinfo {author} {\bibfnamefont
  {T.}~\bibnamefont {Otsuka}},\ }\href {\doibase 10.3847/1538-4357/833/2/179}
  {\bibfield  {journal} {\bibinfo  {journal} {The Astrophysical Journal}\
  }\textbf {\bibinfo {volume} {833}},\ \bibinfo {pages} {179} (\bibinfo {year}
  {2016})}\BibitemShut {NoStop}%
\bibitem [{\citenamefont {Juodagalvis}\ \emph {et~al.}(2010)\citenamefont
  {Juodagalvis}, \citenamefont {Langanke}, \citenamefont {Hix}, \citenamefont
  {Martínez-Pinedo},\ and\ \citenamefont {Sampaio}}]{JUODAGALVIS2010454}%
  \BibitemOpen
  \bibfield  {author} {\bibinfo {author} {\bibfnamefont {A.}~\bibnamefont
  {Juodagalvis}}, \bibinfo {author} {\bibfnamefont {K.}~\bibnamefont
  {Langanke}}, \bibinfo {author} {\bibfnamefont {W.}~\bibnamefont {Hix}},
  \bibinfo {author} {\bibfnamefont {G.}~\bibnamefont {Martínez-Pinedo}}, \
  and\ \bibinfo {author} {\bibfnamefont {J.}~\bibnamefont {Sampaio}},\ }\href
  {\doibase https://doi.org/10.1016/j.nuclphysa.2010.09.012} {\bibfield
  {journal} {\bibinfo  {journal} {Nuclear Physics A}\ }\textbf {\bibinfo
  {volume} {848}},\ \bibinfo {pages} {454 } (\bibinfo {year}
  {2010})}\BibitemShut {NoStop}%
\bibitem [{\citenamefont {Nabi}\ and\ \citenamefont
  {Klapdor-Kleingrothaus}(2004)}]{NABI2004237}%
  \BibitemOpen
  \bibfield  {author} {\bibinfo {author} {\bibfnamefont {J.-U.}\ \bibnamefont
  {Nabi}}\ and\ \bibinfo {author} {\bibfnamefont {H.~V.}\ \bibnamefont
  {Klapdor-Kleingrothaus}},\ }\href {\doibase
  https://doi.org/10.1016/j.adt.2004.09.002} {\bibfield  {journal} {\bibinfo
  {journal} {Atomic Data and Nuclear Data Tables}\ }\textbf {\bibinfo {volume}
  {88}},\ \bibinfo {pages} {237 } (\bibinfo {year} {2004})}\BibitemShut
  {NoStop}%
\bibitem [{\citenamefont {Langanke}\ and\ \citenamefont
  {Mart\'{\i}nez-Pinedo}(2003)}]{RevModPhys.75.819}%
  \BibitemOpen
  \bibfield  {author} {\bibinfo {author} {\bibfnamefont {K.}~\bibnamefont
  {Langanke}}\ and\ \bibinfo {author} {\bibfnamefont {G.}~\bibnamefont
  {Mart\'{\i}nez-Pinedo}},\ }\href {\doibase 10.1103/RevModPhys.75.819}
  {\bibfield  {journal} {\bibinfo  {journal} {Rev. Mod. Phys.}\ }\textbf
  {\bibinfo {volume} {75}},\ \bibinfo {pages} {819} (\bibinfo {year}
  {2003})}\BibitemShut {NoStop}%
\bibitem [{\citenamefont {Dean}\ \emph {et~al.}(1998)\citenamefont {Dean},
  \citenamefont {Langanke}, \citenamefont {Chatterjee}, \citenamefont {Radha},\
  and\ \citenamefont {Strayer}}]{PhysRevC.58.536}%
  \BibitemOpen
  \bibfield  {author} {\bibinfo {author} {\bibfnamefont {D.~J.}\ \bibnamefont
  {Dean}}, \bibinfo {author} {\bibfnamefont {K.}~\bibnamefont {Langanke}},
  \bibinfo {author} {\bibfnamefont {L.}~\bibnamefont {Chatterjee}}, \bibinfo
  {author} {\bibfnamefont {P.~B.}\ \bibnamefont {Radha}}, \ and\ \bibinfo
  {author} {\bibfnamefont {M.~R.}\ \bibnamefont {Strayer}},\ }\href {\doibase
  10.1103/PhysRevC.58.536} {\bibfield  {journal} {\bibinfo  {journal} {Phys.
  Rev. C}\ }\textbf {\bibinfo {volume} {58}},\ \bibinfo {pages} {536} (\bibinfo
  {year} {1998})}\BibitemShut {NoStop}%
\bibitem [{\citenamefont {Cooperstein}\ and\ \citenamefont
  {Wambach}(1984)}]{COOPERSTEIN1984591}%
  \BibitemOpen
  \bibfield  {author} {\bibinfo {author} {\bibfnamefont {J.}~\bibnamefont
  {Cooperstein}}\ and\ \bibinfo {author} {\bibfnamefont {J.}~\bibnamefont
  {Wambach}},\ }\href {\doibase https://doi.org/10.1016/0375-9474(84)90673-0}
  {\bibfield  {journal} {\bibinfo  {journal} {Nuclear Physics A}\ }\textbf
  {\bibinfo {volume} {420}},\ \bibinfo {pages} {591 } (\bibinfo {year}
  {1984})}\BibitemShut {NoStop}%
\bibitem [{\citenamefont {Langanke}\ \emph {et~al.}(2001)\citenamefont
  {Langanke}, \citenamefont {Kolbe},\ and\ \citenamefont
  {Dean}}]{PhysRevC.63.032801}%
  \BibitemOpen
  \bibfield  {author} {\bibinfo {author} {\bibfnamefont {K.}~\bibnamefont
  {Langanke}}, \bibinfo {author} {\bibfnamefont {E.}~\bibnamefont {Kolbe}}, \
  and\ \bibinfo {author} {\bibfnamefont {D.~J.}\ \bibnamefont {Dean}},\ }\href
  {\doibase 10.1103/PhysRevC.63.032801} {\bibfield  {journal} {\bibinfo
  {journal} {Phys. Rev. C}\ }\textbf {\bibinfo {volume} {63}},\ \bibinfo
  {pages} {032801(R)} (\bibinfo {year} {2001})}\BibitemShut {NoStop}%
\bibitem [{\citenamefont {{Fantina, A. F.}}\ \emph {et~al.}(2014)\citenamefont
  {{Fantina, A. F.}}, \citenamefont {{Khan, E.}}, \citenamefont {{Col\`o, G.}},
  \citenamefont {{Paar, N.}},\ and\ \citenamefont {{Vretenar, D.}}}]{refId0}%
  \BibitemOpen
  \bibfield  {author} {\bibinfo {author} {\bibnamefont {{Fantina, A. F.}}},
  \bibinfo {author} {\bibnamefont {{Khan, E.}}}, \bibinfo {author}
  {\bibnamefont {{Col\`o, G.}}}, \bibinfo {author} {\bibnamefont {{Paar, N.}}},
  \ and\ \bibinfo {author} {\bibnamefont {{Vretenar, D.}}},\ }\href {\doibase
  10.1051/epjconf/20146602035} {\bibfield  {journal} {\bibinfo  {journal} {EPJ
  Web of Conferences}\ }\textbf {\bibinfo {volume} {66}},\ \bibinfo {pages}
  {02035} (\bibinfo {year} {2014})}\BibitemShut {NoStop}%
\bibitem [{\citenamefont {Paar}\ \emph {et~al.}(2009)\citenamefont {Paar},
  \citenamefont {Col\`o}, \citenamefont {Khan},\ and\ \citenamefont
  {Vretenar}}]{PhysRevC.80.055801}%
  \BibitemOpen
  \bibfield  {author} {\bibinfo {author} {\bibfnamefont {N.}~\bibnamefont
  {Paar}}, \bibinfo {author} {\bibfnamefont {G.}~\bibnamefont {Col\`o}},
  \bibinfo {author} {\bibfnamefont {E.}~\bibnamefont {Khan}}, \ and\ \bibinfo
  {author} {\bibfnamefont {D.}~\bibnamefont {Vretenar}},\ }\href {\doibase
  10.1103/PhysRevC.80.055801} {\bibfield  {journal} {\bibinfo  {journal} {Phys.
  Rev. C}\ }\textbf {\bibinfo {volume} {80}},\ \bibinfo {pages} {055801}
  (\bibinfo {year} {2009})}\BibitemShut {NoStop}%
\bibitem [{\citenamefont {Fantina}\ \emph {et~al.}(2012)\citenamefont
  {Fantina}, \citenamefont {Khan}, \citenamefont {Col\`o}, \citenamefont
  {Paar},\ and\ \citenamefont {Vretenar}}]{PhysRevC.86.035805}%
  \BibitemOpen
  \bibfield  {author} {\bibinfo {author} {\bibfnamefont {A.~F.}\ \bibnamefont
  {Fantina}}, \bibinfo {author} {\bibfnamefont {E.}~\bibnamefont {Khan}},
  \bibinfo {author} {\bibfnamefont {G.}~\bibnamefont {Col\`o}}, \bibinfo
  {author} {\bibfnamefont {N.}~\bibnamefont {Paar}}, \ and\ \bibinfo {author}
  {\bibfnamefont {D.}~\bibnamefont {Vretenar}},\ }\href {\doibase
  10.1103/PhysRevC.86.035805} {\bibfield  {journal} {\bibinfo  {journal} {Phys.
  Rev. C}\ }\textbf {\bibinfo {volume} {86}},\ \bibinfo {pages} {035805}
  (\bibinfo {year} {2012})}\BibitemShut {NoStop}%
\bibitem [{\citenamefont {Niu}\ \emph {et~al.}(2011)\citenamefont {Niu},
  \citenamefont {Paar}, \citenamefont {Vretenar},\ and\ \citenamefont
  {Meng}}]{PhysRevC.83.045807}%
  \BibitemOpen
  \bibfield  {author} {\bibinfo {author} {\bibfnamefont {Y.~F.}\ \bibnamefont
  {Niu}}, \bibinfo {author} {\bibfnamefont {N.}~\bibnamefont {Paar}}, \bibinfo
  {author} {\bibfnamefont {D.}~\bibnamefont {Vretenar}}, \ and\ \bibinfo
  {author} {\bibfnamefont {J.}~\bibnamefont {Meng}},\ }\href {\doibase
  10.1103/PhysRevC.83.045807} {\bibfield  {journal} {\bibinfo  {journal} {Phys.
  Rev. C}\ }\textbf {\bibinfo {volume} {83}},\ \bibinfo {pages} {045807}
  (\bibinfo {year} {2011})}\BibitemShut {NoStop}%
\bibitem [{\citenamefont {Dzhioev}\ \emph {et~al.}(2010)\citenamefont
  {Dzhioev}, \citenamefont {Vdovin}, \citenamefont {Ponomarev}, \citenamefont
  {Wambach}, \citenamefont {Langanke},\ and\ \citenamefont
  {Mart\'{\i}nez-Pinedo}}]{PhysRevC.81.015804}%
  \BibitemOpen
  \bibfield  {author} {\bibinfo {author} {\bibfnamefont {A.~A.}\ \bibnamefont
  {Dzhioev}}, \bibinfo {author} {\bibfnamefont {A.~I.}\ \bibnamefont {Vdovin}},
  \bibinfo {author} {\bibfnamefont {V.~Y.}\ \bibnamefont {Ponomarev}}, \bibinfo
  {author} {\bibfnamefont {J.}~\bibnamefont {Wambach}}, \bibinfo {author}
  {\bibfnamefont {K.}~\bibnamefont {Langanke}}, \ and\ \bibinfo {author}
  {\bibfnamefont {G.}~\bibnamefont {Mart\'{\i}nez-Pinedo}},\ }\href {\doibase
  10.1103/PhysRevC.81.015804} {\bibfield  {journal} {\bibinfo  {journal} {Phys.
  Rev. C}\ }\textbf {\bibinfo {volume} {81}},\ \bibinfo {pages} {015804}
  (\bibinfo {year} {2010})}\BibitemShut {NoStop}%
\bibitem [{\citenamefont {Dzhioev}\ \emph {et~al.}(2019)\citenamefont
  {Dzhioev}, \citenamefont {Vdovin},\ and\ \citenamefont
  {Stoyanov}}]{PhysRevC.100.025801}%
  \BibitemOpen
  \bibfield  {author} {\bibinfo {author} {\bibfnamefont {A.~A.}\ \bibnamefont
  {Dzhioev}}, \bibinfo {author} {\bibfnamefont {A.~I.}\ \bibnamefont {Vdovin}},
  \ and\ \bibinfo {author} {\bibfnamefont {C.}~\bibnamefont {Stoyanov}},\
  }\href {\doibase 10.1103/PhysRevC.100.025801} {\bibfield  {journal} {\bibinfo
   {journal} {Phys. Rev. C}\ }\textbf {\bibinfo {volume} {100}},\ \bibinfo
  {pages} {025801} (\bibinfo {year} {2019})}\BibitemShut {NoStop}%
\bibitem [{\citenamefont {Dzhioev}\ \emph {et~al.}(2020)\citenamefont
  {Dzhioev}, \citenamefont {Langanke}, \citenamefont {Mart\'{\i}nez-Pinedo},
  \citenamefont {Vdovin},\ and\ \citenamefont
  {Stoyanov}}]{PhysRevC.101.025805}%
  \BibitemOpen
  \bibfield  {author} {\bibinfo {author} {\bibfnamefont {A.~A.}\ \bibnamefont
  {Dzhioev}}, \bibinfo {author} {\bibfnamefont {K.}~\bibnamefont {Langanke}},
  \bibinfo {author} {\bibfnamefont {G.}~\bibnamefont {Mart\'{\i}nez-Pinedo}},
  \bibinfo {author} {\bibfnamefont {A.~I.}\ \bibnamefont {Vdovin}}, \ and\
  \bibinfo {author} {\bibfnamefont {C.}~\bibnamefont {Stoyanov}},\ }\href
  {\doibase 10.1103/PhysRevC.101.025805} {\bibfield  {journal} {\bibinfo
  {journal} {Phys. Rev. C}\ }\textbf {\bibinfo {volume} {101}},\ \bibinfo
  {pages} {025805} (\bibinfo {year} {2020})}\BibitemShut {NoStop}%
\bibitem [{\citenamefont {Litvinova}\ \emph {et~al.}(2020)\citenamefont
  {Litvinova}, \citenamefont {Robin},\ and\ \citenamefont {Wibowo}}]{lit18}%
  \BibitemOpen
  \bibfield  {author} {\bibinfo {author} {\bibfnamefont {E.}~\bibnamefont
  {Litvinova}}, \bibinfo {author} {\bibfnamefont {C.}~\bibnamefont {Robin}}, \
  and\ \bibinfo {author} {\bibfnamefont {H.}~\bibnamefont {Wibowo}},\ }\href
  {\doibase 10.1016/j.physletb.2019.135134} {\bibfield  {journal} {\bibinfo
  {journal} {Physics Letters B}\ }\textbf {\bibinfo {volume} {800}},\ \bibinfo
  {pages} {135134} (\bibinfo {year} {2020})}\BibitemShut {NoStop}%
\bibitem [{\citenamefont {Nikšić}\ \emph {et~al.}(2011)\citenamefont
  {Nikšić}, \citenamefont {Vretenar},\ and\ \citenamefont
  {Ring}}]{NIKSIC2011519}%
  \BibitemOpen
  \bibfield  {author} {\bibinfo {author} {\bibfnamefont {T.}~\bibnamefont
  {Nikšić}}, \bibinfo {author} {\bibfnamefont {D.}~\bibnamefont {Vretenar}},
  \ and\ \bibinfo {author} {\bibfnamefont {P.}~\bibnamefont {Ring}},\ }\href
  {\doibase https://doi.org/10.1016/j.ppnp.2011.01.055} {\bibfield  {journal}
  {\bibinfo  {journal} {Progress in Particle and Nuclear Physics}\ }\textbf
  {\bibinfo {volume} {66}},\ \bibinfo {pages} {519 } (\bibinfo {year}
  {2011})}\BibitemShut {NoStop}%
\bibitem [{\citenamefont {Nik{\v s}i{\'c}}\ \emph {et~al.}(2014)\citenamefont
  {Nik{\v s}i{\'c}}, \citenamefont {Paar}, \citenamefont {Vretenar},\ and\
  \citenamefont {Ring}}]{NIKSIC20141808}%
  \BibitemOpen
  \bibfield  {author} {\bibinfo {author} {\bibfnamefont {T.}~\bibnamefont
  {Nik{\v s}i{\'c}}}, \bibinfo {author} {\bibfnamefont {N.}~\bibnamefont
  {Paar}}, \bibinfo {author} {\bibfnamefont {D.}~\bibnamefont {Vretenar}}, \
  and\ \bibinfo {author} {\bibfnamefont {P.}~\bibnamefont {Ring}},\ }\href
  {\doibase https://doi.org/10.1016/j.cpc.2014.02.027} {\bibfield  {journal}
  {\bibinfo  {journal} {Computer Physics Communications}\ }\textbf {\bibinfo
  {volume} {185}},\ \bibinfo {pages} {1808 } (\bibinfo {year}
  {2014})}\BibitemShut {NoStop}%
\bibitem [{\citenamefont {Y\"uksel}\ \emph {et~al.}(2020)\citenamefont
  {Y\"uksel}, \citenamefont {Paar}, \citenamefont {Col\`o}, \citenamefont
  {Khan},\ and\ \citenamefont {Niu}}]{yksel2019gamowteller}%
  \BibitemOpen
  \bibfield  {author} {\bibinfo {author} {\bibfnamefont {E.}~\bibnamefont
  {Y\"uksel}}, \bibinfo {author} {\bibfnamefont {N.}~\bibnamefont {Paar}},
  \bibinfo {author} {\bibfnamefont {G.}~\bibnamefont {Col\`o}}, \bibinfo
  {author} {\bibfnamefont {E.}~\bibnamefont {Khan}}, \ and\ \bibinfo {author}
  {\bibfnamefont {Y.~F.}\ \bibnamefont {Niu}},\ }\href {\doibase
  10.1103/PhysRevC.101.044305} {\bibfield  {journal} {\bibinfo  {journal}
  {Phys. Rev. C}\ }\textbf {\bibinfo {volume} {101}},\ \bibinfo {pages}
  {044305} (\bibinfo {year} {2020})}\BibitemShut {NoStop}%
\bibitem [{\citenamefont {Y{\"u}ksel}\ \emph {et~al.}(2019)\citenamefont
  {Y{\"u}ksel}, \citenamefont {Colò}, \citenamefont {Khan},\ and\
  \citenamefont {Niu}}]{yksel2019nuclear}%
  \BibitemOpen
  \bibfield  {author} {\bibinfo {author} {\bibfnamefont {E.}~\bibnamefont
  {Y{\"u}ksel}}, \bibinfo {author} {\bibfnamefont {G.}~\bibnamefont {Colò}},
  \bibinfo {author} {\bibfnamefont {E.}~\bibnamefont {Khan}}, \ and\ \bibinfo
  {author} {\bibfnamefont {Y.}~\bibnamefont {Niu}},\ }\href {\doibase
  10.1140/epja/i2019-12918-8} {\bibfield  {journal} {\bibinfo  {journal} {Eur.
  Phys. J. A}\ }\textbf {\bibinfo {volume} {55}},\ \bibinfo {pages} {230}
  (\bibinfo {year} {2019})}\BibitemShut {NoStop}%
\bibitem [{\citenamefont {Ring}(1996)}]{RING1996193}%
  \BibitemOpen
  \bibfield  {author} {\bibinfo {author} {\bibfnamefont {P.}~\bibnamefont
  {Ring}},\ }\href {\doibase https://doi.org/10.1016/0146-6410(96)00054-3}
  {\bibfield  {journal} {\bibinfo  {journal} {Progress in Particle and Nuclear
  Physics}\ }\textbf {\bibinfo {volume} {37}},\ \bibinfo {pages} {193 }
  (\bibinfo {year} {1996})}\BibitemShut {NoStop}%
\bibitem [{\citenamefont {Gambhir}\ \emph {et~al.}(1990)\citenamefont
  {Gambhir}, \citenamefont {Ring},\ and\ \citenamefont
  {Thimet}}]{GAMBHIR1990132}%
  \BibitemOpen
  \bibfield  {author} {\bibinfo {author} {\bibfnamefont {Y.}~\bibnamefont
  {Gambhir}}, \bibinfo {author} {\bibfnamefont {P.}~\bibnamefont {Ring}}, \
  and\ \bibinfo {author} {\bibfnamefont {A.}~\bibnamefont {Thimet}},\ }\href
  {\doibase https://doi.org/10.1016/0003-4916(90)90330-Q} {\bibfield  {journal}
  {\bibinfo  {journal} {Annals of Physics}\ }\textbf {\bibinfo {volume}
  {198}},\ \bibinfo {pages} {132 } (\bibinfo {year} {1990})}\BibitemShut
  {NoStop}%
\bibitem [{\citenamefont {Reinhard}(1989)}]{Reinhard_1989}%
  \BibitemOpen
  \bibfield  {author} {\bibinfo {author} {\bibfnamefont {P.~G.}\ \bibnamefont
  {Reinhard}},\ }\href {\doibase 10.1088/0034-4885/52/4/002} {\bibfield
  {journal} {\bibinfo  {journal} {Reports on Progress in Physics}\ }\textbf
  {\bibinfo {volume} {52}},\ \bibinfo {pages} {439} (\bibinfo {year}
  {1989})}\BibitemShut {NoStop}%
\bibitem [{\citenamefont {Nik\ifmmode \check{s}\else
  \v{s}\fi{}i\ifmmode~\acute{c}\else \'{c}\fi{}}\ \emph
  {et~al.}(2002)\citenamefont {Nik\ifmmode \check{s}\else
  \v{s}\fi{}i\ifmmode~\acute{c}\else \'{c}\fi{}}, \citenamefont {Vretenar},
  \citenamefont {Finelli},\ and\ \citenamefont {Ring}}]{PhysRevC.66.024306}%
  \BibitemOpen
  \bibfield  {author} {\bibinfo {author} {\bibfnamefont {T.}~\bibnamefont
  {Nik\ifmmode \check{s}\else \v{s}\fi{}i\ifmmode~\acute{c}\else \'{c}\fi{}}},
  \bibinfo {author} {\bibfnamefont {D.}~\bibnamefont {Vretenar}}, \bibinfo
  {author} {\bibfnamefont {P.}~\bibnamefont {Finelli}}, \ and\ \bibinfo
  {author} {\bibfnamefont {P.}~\bibnamefont {Ring}},\ }\href {\doibase
  10.1103/PhysRevC.66.024306} {\bibfield  {journal} {\bibinfo  {journal} {Phys.
  Rev. C}\ }\textbf {\bibinfo {volume} {66}},\ \bibinfo {pages} {024306}
  (\bibinfo {year} {2002})}\BibitemShut {NoStop}%
\bibitem [{\citenamefont {Lalazissis}\ \emph {et~al.}(2005)\citenamefont
  {Lalazissis}, \citenamefont {Nik\ifmmode \check{s}\else
  \v{s}\fi{}i\ifmmode~\acute{c}\else \'{c}\fi{}}, \citenamefont {Vretenar},\
  and\ \citenamefont {Ring}}]{PhysRevC.71.024312}%
  \BibitemOpen
  \bibfield  {author} {\bibinfo {author} {\bibfnamefont {G.~A.}\ \bibnamefont
  {Lalazissis}}, \bibinfo {author} {\bibfnamefont {T.}~\bibnamefont
  {Nik\ifmmode \check{s}\else \v{s}\fi{}i\ifmmode~\acute{c}\else \'{c}\fi{}}},
  \bibinfo {author} {\bibfnamefont {D.}~\bibnamefont {Vretenar}}, \ and\
  \bibinfo {author} {\bibfnamefont {P.}~\bibnamefont {Ring}},\ }\href {\doibase
  10.1103/PhysRevC.71.024312} {\bibfield  {journal} {\bibinfo  {journal} {Phys.
  Rev. C}\ }\textbf {\bibinfo {volume} {71}},\ \bibinfo {pages} {024312}
  (\bibinfo {year} {2005})}\BibitemShut {NoStop}%
\bibitem [{\citenamefont {Goodman}(1981)}]{GOODMAN198130}%
  \BibitemOpen
  \bibfield  {author} {\bibinfo {author} {\bibfnamefont {A.~L.}\ \bibnamefont
  {Goodman}},\ }\href {\doibase https://doi.org/10.1016/0375-9474(81)90557-1}
  {\bibfield  {journal} {\bibinfo  {journal} {Nuclear Physics A}\ }\textbf
  {\bibinfo {volume} {352}},\ \bibinfo {pages} {30 } (\bibinfo {year}
  {1981})}\BibitemShut {NoStop}%
\bibitem [{\citenamefont {{Y\"uksel}}\ \emph {et~al.}(2014)\citenamefont
  {{Y\"uksel}}, \citenamefont {{E. Khan}}, \citenamefont {{K. Bozkurt}},\ and\
  \citenamefont {{G. Col\`o}}}]{yuksel2014effect}%
  \BibitemOpen
  \bibfield  {author} {\bibinfo {author} {\bibfnamefont {E.}~\bibnamefont
  {{Y\"uksel}}}, \bibinfo {author} {\bibnamefont {{E. Khan}}}, \bibinfo
  {author} {\bibnamefont {{K. Bozkurt}}}, \ and\ \bibinfo {author}
  {\bibnamefont {{G. Col\`o}}},\ }\href {\doibase 10.1140/epja/i2014-14160-4}
  {\bibfield  {journal} {\bibinfo  {journal} {Eur. Phys. J. A}\ }\textbf
  {\bibinfo {volume} {50}},\ \bibinfo {pages} {160} (\bibinfo {year}
  {2014})}\BibitemShut {NoStop}%
\bibitem [{\citenamefont {Ring}\ and\ \citenamefont
  {Schuck}(2004)}]{ring2004nuclear}%
  \BibitemOpen
  \bibfield  {author} {\bibinfo {author} {\bibfnamefont {P.}~\bibnamefont
  {Ring}}\ and\ \bibinfo {author} {\bibfnamefont {P.}~\bibnamefont {Schuck}},\
  }\href {https://books.google.hr/books?id=PTynSM-nMA8C} {\emph {\bibinfo
  {title} {The Nuclear Many-Body Problem}}},\ Physics and astronomy online
  library\ (\bibinfo  {publisher} {Springer},\ \bibinfo {year}
  {2004})\BibitemShut {NoStop}%
\bibitem [{\citenamefont {Bender}\ \emph {et~al.}(2000)\citenamefont {Bender},
  \citenamefont {Rutz}, \citenamefont {Reinhard},\ and\ \citenamefont
  {Maruhn}}]{Bender2000}%
  \BibitemOpen
  \bibfield  {author} {\bibinfo {author} {\bibfnamefont {M.}~\bibnamefont
  {Bender}}, \bibinfo {author} {\bibfnamefont {K.}~\bibnamefont {Rutz}},
  \bibinfo {author} {\bibfnamefont {P.~G.}\ \bibnamefont {Reinhard}}, \ and\
  \bibinfo {author} {\bibfnamefont {J.~A.}\ \bibnamefont {Maruhn}},\ }\href
  {\doibase 10.1007/s10050-000-4504-z} {\bibfield  {journal} {\bibinfo
  {journal} {The European Physical Journal A}\ }\textbf {\bibinfo {volume}
  {8}},\ \bibinfo {pages} {59} (\bibinfo {year} {2000})}\BibitemShut {NoStop}%
\bibitem [{\citenamefont {M{\"o}ller}\ and\ \citenamefont
  {Nix}(1992)}]{MOLLER199220}%
  \BibitemOpen
  \bibfield  {author} {\bibinfo {author} {\bibfnamefont {P.}~\bibnamefont
  {M{\"o}ller}}\ and\ \bibinfo {author} {\bibfnamefont {J.}~\bibnamefont
  {Nix}},\ }\href {\doibase https://doi.org/10.1016/0375-9474(92)90244-E}
  {\bibfield  {journal} {\bibinfo  {journal} {Nuclear Physics A}\ }\textbf
  {\bibinfo {volume} {536}},\ \bibinfo {pages} {20 } (\bibinfo {year}
  {1992})}\BibitemShut {NoStop}%
\bibitem [{\citenamefont {Paar}\ \emph {et~al.}(2004)\citenamefont {Paar},
  \citenamefont {Nik\ifmmode \check{s}\else \v{s}\fi{}i\ifmmode~\acute{c}\else
  \'{c}\fi{}}, \citenamefont {Vretenar},\ and\ \citenamefont
  {Ring}}]{PhysRevC.69.054303}%
  \BibitemOpen
  \bibfield  {author} {\bibinfo {author} {\bibfnamefont {N.}~\bibnamefont
  {Paar}}, \bibinfo {author} {\bibfnamefont {T.}~\bibnamefont {Nik\ifmmode
  \check{s}\else \v{s}\fi{}i\ifmmode~\acute{c}\else \'{c}\fi{}}}, \bibinfo
  {author} {\bibfnamefont {D.}~\bibnamefont {Vretenar}}, \ and\ \bibinfo
  {author} {\bibfnamefont {P.}~\bibnamefont {Ring}},\ }\href {\doibase
  10.1103/PhysRevC.69.054303} {\bibfield  {journal} {\bibinfo  {journal} {Phys.
  Rev. C}\ }\textbf {\bibinfo {volume} {69}},\ \bibinfo {pages} {054303}
  (\bibinfo {year} {2004})}\BibitemShut {NoStop}%
\bibitem [{\citenamefont {Niu}\ \emph {et~al.}(2013)\citenamefont {Niu},
  \citenamefont {Niu}, \citenamefont {Paar}, \citenamefont {Vretenar},
  \citenamefont {Wang}, \citenamefont {Bai},\ and\ \citenamefont
  {Meng}}]{PhysRevC.88.034308}%
  \BibitemOpen
  \bibfield  {author} {\bibinfo {author} {\bibfnamefont {Y.~F.}\ \bibnamefont
  {Niu}}, \bibinfo {author} {\bibfnamefont {Z.~M.}\ \bibnamefont {Niu}},
  \bibinfo {author} {\bibfnamefont {N.}~\bibnamefont {Paar}}, \bibinfo {author}
  {\bibfnamefont {D.}~\bibnamefont {Vretenar}}, \bibinfo {author}
  {\bibfnamefont {G.~H.}\ \bibnamefont {Wang}}, \bibinfo {author}
  {\bibfnamefont {J.~S.}\ \bibnamefont {Bai}}, \ and\ \bibinfo {author}
  {\bibfnamefont {J.}~\bibnamefont {Meng}},\ }\href {\doibase
  10.1103/PhysRevC.88.034308} {\bibfield  {journal} {\bibinfo  {journal} {Phys.
  Rev. C}\ }\textbf {\bibinfo {volume} {88}},\ \bibinfo {pages} {034308}
  (\bibinfo {year} {2013})}\BibitemShut {NoStop}%
\bibitem [{\citenamefont {Sommermann}(1983)}]{SOMMERMANN1983163}%
  \BibitemOpen
  \bibfield  {author} {\bibinfo {author} {\bibfnamefont {H.}~\bibnamefont
  {Sommermann}},\ }\href {\doibase
  https://doi.org/10.1016/0003-4916(83)90318-4} {\bibfield  {journal} {\bibinfo
   {journal} {Annals of Physics}\ }\textbf {\bibinfo {volume} {151}},\ \bibinfo
  {pages} {163 } (\bibinfo {year} {1983})}\BibitemShut {NoStop}%
\bibitem [{\citenamefont {Y\"uksel}\ \emph {et~al.}(2017)\citenamefont
  {Y\"uksel}, \citenamefont {Col\`o}, \citenamefont {Khan}, \citenamefont
  {Niu},\ and\ \citenamefont {Bozkurt}}]{PhysRevC.96.024303}%
  \BibitemOpen
  \bibfield  {author} {\bibinfo {author} {\bibfnamefont {E.}~\bibnamefont
  {Y\"uksel}}, \bibinfo {author} {\bibfnamefont {G.}~\bibnamefont {Col\`o}},
  \bibinfo {author} {\bibfnamefont {E.}~\bibnamefont {Khan}}, \bibinfo {author}
  {\bibfnamefont {Y.~F.}\ \bibnamefont {Niu}}, \ and\ \bibinfo {author}
  {\bibfnamefont {K.}~\bibnamefont {Bozkurt}},\ }\href {\doibase
  10.1103/PhysRevC.96.024303} {\bibfield  {journal} {\bibinfo  {journal} {Phys.
  Rev. C}\ }\textbf {\bibinfo {volume} {96}},\ \bibinfo {pages} {024303}
  (\bibinfo {year} {2017})}\BibitemShut {NoStop}%
\bibitem [{\citenamefont {Walecka}(1975)}]{walecka1975muon}%
  \BibitemOpen
  \bibfield  {author} {\bibinfo {author} {\bibfnamefont {J.}~\bibnamefont
  {Walecka}},\ }\href@noop {} {\bibfield  {journal} {\bibinfo  {journal}
  {Academis, New York USA}\ }\textbf {\bibinfo {volume} {Vol. II.}} (\bibinfo
  {year} {1975})}\BibitemShut {NoStop}%
\bibitem [{\citenamefont {O'Connell}\ \emph {et~al.}(1972)\citenamefont
  {O'Connell}, \citenamefont {Donnelly},\ and\ \citenamefont
  {Walecka}}]{PhysRevC.6.719}%
  \BibitemOpen
  \bibfield  {author} {\bibinfo {author} {\bibfnamefont {J.~S.}\ \bibnamefont
  {O'Connell}}, \bibinfo {author} {\bibfnamefont {T.~W.}\ \bibnamefont
  {Donnelly}}, \ and\ \bibinfo {author} {\bibfnamefont {J.~D.}\ \bibnamefont
  {Walecka}},\ }\href {\doibase 10.1103/PhysRevC.6.719} {\bibfield  {journal}
  {\bibinfo  {journal} {Phys. Rev. C}\ }\textbf {\bibinfo {volume} {6}},\
  \bibinfo {pages} {719} (\bibinfo {year} {1972})}\BibitemShut {NoStop}%
\bibitem [{\citenamefont {Walecka}(2004)}]{walecka2004theoretical}%
  \BibitemOpen
  \bibfield  {author} {\bibinfo {author} {\bibfnamefont {J.}~\bibnamefont
  {Walecka}},\ }\href {https://books.google.hr/books?id=mfphXc8b-2IC} {\emph
  {\bibinfo {title} {Theoretical Nuclear and Subnuclear Physics}}},\
  Theoretical Nuclear and Subnuclear Physics\ (\bibinfo  {publisher} {Imperial
  College Press},\ \bibinfo {year} {2004})\BibitemShut {NoStop}%
\bibitem [{\citenamefont {Kolbe}\ \emph {et~al.}(2003)\citenamefont {Kolbe},
  \citenamefont {Langanke}, \citenamefont {Mart{\'{\i}}nez-Pinedo},\ and\
  \citenamefont {Vogel}}]{Kolbe_2003}%
  \BibitemOpen
  \bibfield  {author} {\bibinfo {author} {\bibfnamefont {E.}~\bibnamefont
  {Kolbe}}, \bibinfo {author} {\bibfnamefont {K.}~\bibnamefont {Langanke}},
  \bibinfo {author} {\bibfnamefont {G.}~\bibnamefont {Mart{\'{\i}}nez-Pinedo}},
  \ and\ \bibinfo {author} {\bibfnamefont {P.}~\bibnamefont {Vogel}},\ }\href
  {\doibase 10.1088/0954-3899/29/11/010} {\bibfield  {journal} {\bibinfo
  {journal} {Journal of Physics G: Nuclear and Particle Physics}\ }\textbf
  {\bibinfo {volume} {29}},\ \bibinfo {pages} {2569} (\bibinfo {year}
  {2003})}\BibitemShut {NoStop}%
\bibitem [{\citenamefont {Marketin}\ \emph {et~al.}(2009)\citenamefont
  {Marketin}, \citenamefont {Paar}, \citenamefont {Nik\ifmmode \check{s}\else
  \v{s}\fi{}i\ifmmode~\acute{c}\else \'{c}\fi{}},\ and\ \citenamefont
  {Vretenar}}]{PhysRevC.79.054323}%
  \BibitemOpen
  \bibfield  {author} {\bibinfo {author} {\bibfnamefont {T.}~\bibnamefont
  {Marketin}}, \bibinfo {author} {\bibfnamefont {N.}~\bibnamefont {Paar}},
  \bibinfo {author} {\bibfnamefont {T.}~\bibnamefont {Nik\ifmmode
  \check{s}\else \v{s}\fi{}i\ifmmode~\acute{c}\else \'{c}\fi{}}}, \ and\
  \bibinfo {author} {\bibfnamefont {D.}~\bibnamefont {Vretenar}},\ }\href
  {\doibase 10.1103/PhysRevC.79.054323} {\bibfield  {journal} {\bibinfo
  {journal} {Phys. Rev. C}\ }\textbf {\bibinfo {volume} {79}},\ \bibinfo
  {pages} {054323} (\bibinfo {year} {2009})}\BibitemShut {NoStop}%
\bibitem [{\citenamefont {Cole}\ \emph {et~al.}(2012)\citenamefont {Cole},
  \citenamefont {Anderson}, \citenamefont {Zegers}, \citenamefont {Austin},
  \citenamefont {Brown}, \citenamefont {Valdez}, \citenamefont {Gupta},
  \citenamefont {Hitt},\ and\ \citenamefont {Fawwaz}}]{PhysRevC.86.015809}%
  \BibitemOpen
  \bibfield  {author} {\bibinfo {author} {\bibfnamefont {A.~L.}\ \bibnamefont
  {Cole}}, \bibinfo {author} {\bibfnamefont {T.~S.}\ \bibnamefont {Anderson}},
  \bibinfo {author} {\bibfnamefont {R.~G.~T.}\ \bibnamefont {Zegers}}, \bibinfo
  {author} {\bibfnamefont {S.~M.}\ \bibnamefont {Austin}}, \bibinfo {author}
  {\bibfnamefont {B.~A.}\ \bibnamefont {Brown}}, \bibinfo {author}
  {\bibfnamefont {L.}~\bibnamefont {Valdez}}, \bibinfo {author} {\bibfnamefont
  {S.}~\bibnamefont {Gupta}}, \bibinfo {author} {\bibfnamefont {G.~W.}\
  \bibnamefont {Hitt}}, \ and\ \bibinfo {author} {\bibfnamefont
  {O.}~\bibnamefont {Fawwaz}},\ }\href {\doibase 10.1103/PhysRevC.86.015809}
  {\bibfield  {journal} {\bibinfo  {journal} {Phys. Rev. C}\ }\textbf {\bibinfo
  {volume} {86}},\ \bibinfo {pages} {015809} (\bibinfo {year}
  {2012})}\BibitemShut {NoStop}%
\bibitem [{\citenamefont {Heger}\ \emph {et~al.}(2001)\citenamefont {Heger},
  \citenamefont {Woosley}, \citenamefont {Martinez-Pinedo},\ and\ \citenamefont
  {Langanke}}]{Heger_2001}%
  \BibitemOpen
  \bibfield  {author} {\bibinfo {author} {\bibfnamefont {A.}~\bibnamefont
  {Heger}}, \bibinfo {author} {\bibfnamefont {S.~E.}\ \bibnamefont {Woosley}},
  \bibinfo {author} {\bibfnamefont {G.}~\bibnamefont {Martinez-Pinedo}}, \ and\
  \bibinfo {author} {\bibfnamefont {K.}~\bibnamefont {Langanke}},\ }\href
  {\doibase 10.1086/324092} {\bibfield  {journal} {\bibinfo  {journal} {The
  Astrophysical Journal}\ }\textbf {\bibinfo {volume} {560}},\ \bibinfo {pages}
  {307} (\bibinfo {year} {2001})}\BibitemShut {NoStop}%
\bibitem [{\citenamefont {Niu}\ \emph {et~al.}(2017)\citenamefont {Niu},
  \citenamefont {Niu}, \citenamefont {Liang}, \citenamefont {Long},\ and\
  \citenamefont {Meng}}]{PhysRevC.95.044301}%
  \BibitemOpen
  \bibfield  {author} {\bibinfo {author} {\bibfnamefont {Z.~M.}\ \bibnamefont
  {Niu}}, \bibinfo {author} {\bibfnamefont {Y.~F.}\ \bibnamefont {Niu}},
  \bibinfo {author} {\bibfnamefont {H.~Z.}\ \bibnamefont {Liang}}, \bibinfo
  {author} {\bibfnamefont {W.~H.}\ \bibnamefont {Long}}, \ and\ \bibinfo
  {author} {\bibfnamefont {J.}~\bibnamefont {Meng}},\ }\href {\doibase
  10.1103/PhysRevC.95.044301} {\bibfield  {journal} {\bibinfo  {journal} {Phys.
  Rev. C}\ }\textbf {\bibinfo {volume} {95}},\ \bibinfo {pages} {044301}
  (\bibinfo {year} {2017})}\BibitemShut {NoStop}%
\bibitem [{\citenamefont {Fracasso}\ and\ \citenamefont
  {Col\`o}(2007)}]{PhysRevC.76.044307}%
  \BibitemOpen
  \bibfield  {author} {\bibinfo {author} {\bibfnamefont {S.}~\bibnamefont
  {Fracasso}}\ and\ \bibinfo {author} {\bibfnamefont {G.}~\bibnamefont
  {Col\`o}},\ }\href {\doibase 10.1103/PhysRevC.76.044307} {\bibfield
  {journal} {\bibinfo  {journal} {Phys. Rev. C}\ }\textbf {\bibinfo {volume}
  {76}},\ \bibinfo {pages} {044307} (\bibinfo {year} {2007})}\BibitemShut
  {NoStop}%
\bibitem [{\citenamefont {Bai}\ \emph {et~al.}(2014)\citenamefont {Bai},
  \citenamefont {Sagawa}, \citenamefont {Col\`o}, \citenamefont {Fujita},
  \citenamefont {Zhang}, \citenamefont {Zhang},\ and\ \citenamefont
  {Xu}}]{PhysRevC.90.054335}%
  \BibitemOpen
  \bibfield  {author} {\bibinfo {author} {\bibfnamefont {C.~L.}\ \bibnamefont
  {Bai}}, \bibinfo {author} {\bibfnamefont {H.}~\bibnamefont {Sagawa}},
  \bibinfo {author} {\bibfnamefont {G.}~\bibnamefont {Col\`o}}, \bibinfo
  {author} {\bibfnamefont {Y.}~\bibnamefont {Fujita}}, \bibinfo {author}
  {\bibfnamefont {H.~Q.}\ \bibnamefont {Zhang}}, \bibinfo {author}
  {\bibfnamefont {X.~Z.}\ \bibnamefont {Zhang}}, \ and\ \bibinfo {author}
  {\bibfnamefont {F.~R.}\ \bibnamefont {Xu}},\ }\href {\doibase
  10.1103/PhysRevC.90.054335} {\bibfield  {journal} {\bibinfo  {journal} {Phys.
  Rev. C}\ }\textbf {\bibinfo {volume} {90}},\ \bibinfo {pages} {054335}
  (\bibinfo {year} {2014})}\BibitemShut {NoStop}%
\bibitem [{\citenamefont {Bai}\ \emph {et~al.}(2013)\citenamefont {Bai},
  \citenamefont {Sagawa}, \citenamefont {Sasano}, \citenamefont {Uesaka},
  \citenamefont {Hagino}, \citenamefont {Zhang}, \citenamefont {Zhang},\ and\
  \citenamefont {Xu}}]{BAI2013116}%
  \BibitemOpen
  \bibfield  {author} {\bibinfo {author} {\bibfnamefont {C.~L.}\ \bibnamefont
  {Bai}}, \bibinfo {author} {\bibfnamefont {H.}~\bibnamefont {Sagawa}},
  \bibinfo {author} {\bibfnamefont {M.}~\bibnamefont {Sasano}}, \bibinfo
  {author} {\bibfnamefont {T.}~\bibnamefont {Uesaka}}, \bibinfo {author}
  {\bibfnamefont {K.}~\bibnamefont {Hagino}}, \bibinfo {author} {\bibfnamefont
  {H.~Q.}\ \bibnamefont {Zhang}}, \bibinfo {author} {\bibfnamefont {X.~Z.}\
  \bibnamefont {Zhang}}, \ and\ \bibinfo {author} {\bibfnamefont {F.~R.}\
  \bibnamefont {Xu}},\ }\href {\doibase 10.1016/j.physletb.2012.12.060}
  {\bibfield  {journal} {\bibinfo  {journal} {Physics Letters B}\ }\textbf
  {\bibinfo {volume} {719}},\ \bibinfo {pages} {116} (\bibinfo {year}
  {2013})}\BibitemShut {NoStop}%
\bibitem [{\citenamefont {Engel}\ \emph {et~al.}(1999)\citenamefont {Engel},
  \citenamefont {Bender}, \citenamefont {Dobaczewski}, \citenamefont
  {Nazarewicz},\ and\ \citenamefont {Surman}}]{PhysRevC.60.014302}%
  \BibitemOpen
  \bibfield  {author} {\bibinfo {author} {\bibfnamefont {J.}~\bibnamefont
  {Engel}}, \bibinfo {author} {\bibfnamefont {M.}~\bibnamefont {Bender}},
  \bibinfo {author} {\bibfnamefont {J.}~\bibnamefont {Dobaczewski}}, \bibinfo
  {author} {\bibfnamefont {W.}~\bibnamefont {Nazarewicz}}, \ and\ \bibinfo
  {author} {\bibfnamefont {R.}~\bibnamefont {Surman}},\ }\href {\doibase
  10.1103/PhysRevC.60.014302} {\bibfield  {journal} {\bibinfo  {journal} {Phys.
  Rev. C}\ }\textbf {\bibinfo {volume} {60}},\ \bibinfo {pages} {014302}
  (\bibinfo {year} {1999})}\BibitemShut {NoStop}%
\bibitem [{\citenamefont {Sagawa}\ \emph {et~al.}(2016)\citenamefont {Sagawa},
  \citenamefont {Bai},\ and\ \citenamefont {Col{\`{o}}}}]{Sagawa_2016}%
  \BibitemOpen
  \bibfield  {author} {\bibinfo {author} {\bibfnamefont {H.}~\bibnamefont
  {Sagawa}}, \bibinfo {author} {\bibfnamefont {C.~L.}\ \bibnamefont {Bai}}, \
  and\ \bibinfo {author} {\bibfnamefont {G.}~\bibnamefont {Col{\`{o}}}},\
  }\href {\doibase 10.1088/0031-8949/91/8/083011} {\bibfield  {journal}
  {\bibinfo  {journal} {Physica Scripta}\ }\textbf {\bibinfo {volume} {91}},\
  \bibinfo {pages} {083011} (\bibinfo {year} {2016})}\BibitemShut {NoStop}%
\bibitem [{\citenamefont {Niu}\ \emph {et~al.}(2018)\citenamefont {Niu},
  \citenamefont {Niu}, \citenamefont {Col\`{o}},\ and\ \citenamefont
  {Vigezzi}}]{NIU2018325}%
  \BibitemOpen
  \bibfield  {author} {\bibinfo {author} {\bibfnamefont {Y.~F.}\ \bibnamefont
  {Niu}}, \bibinfo {author} {\bibfnamefont {Z.~M.}\ \bibnamefont {Niu}},
  \bibinfo {author} {\bibfnamefont {G.}~\bibnamefont {Col\`{o}}}, \ and\
  \bibinfo {author} {\bibfnamefont {E.}~\bibnamefont {Vigezzi}},\ }\href
  {\doibase 10.1016/j.physletb.2018.02.061} {\bibfield  {journal} {\bibinfo
  {journal} {Physics Letters B}\ }\textbf {\bibinfo {volume} {780}},\ \bibinfo
  {pages} {325} (\bibinfo {year} {2018})}\BibitemShut {NoStop}%
\bibitem [{\citenamefont {Sullivan}\ \emph {et~al.}(2015)\citenamefont
  {Sullivan}, \citenamefont {O'Connor}, \citenamefont {Zegers}, \citenamefont
  {Grubb},\ and\ \citenamefont {Austin}}]{Sullivan_2015}%
  \BibitemOpen
  \bibfield  {author} {\bibinfo {author} {\bibfnamefont {C.}~\bibnamefont
  {Sullivan}}, \bibinfo {author} {\bibfnamefont {E.}~\bibnamefont {O'Connor}},
  \bibinfo {author} {\bibfnamefont {R.~G.~T.}\ \bibnamefont {Zegers}}, \bibinfo
  {author} {\bibfnamefont {T.}~\bibnamefont {Grubb}}, \ and\ \bibinfo {author}
  {\bibfnamefont {S.~M.}\ \bibnamefont {Austin}},\ }\href {\doibase
  10.3847/0004-637x/816/1/44} {\bibfield  {journal} {\bibinfo  {journal} {The
  Astrophysical Journal}\ }\textbf {\bibinfo {volume} {816}},\ \bibinfo {pages}
  {44} (\bibinfo {year} {2015})}\BibitemShut {NoStop}%
\bibitem [{\citenamefont {Suzuki}(2019)}]{SUZUKI2019}%
  \BibitemOpen
  \bibfield  {author} {\bibinfo {author} {\bibfnamefont {T.}~\bibnamefont
  {Suzuki}},\ }\href@noop {} {\bibfield  {journal} {\bibinfo  {journal}
  {Private communications}\ } (\bibinfo {year} {2019})}\BibitemShut {NoStop}%
\bibitem [{\citenamefont {Langanke}\ and\ \citenamefont
  {Martinez-Pinedo}(2000)}]{LANGANKE2000481}%
  \BibitemOpen
  \bibfield  {author} {\bibinfo {author} {\bibfnamefont {K.}~\bibnamefont
  {Langanke}}\ and\ \bibinfo {author} {\bibfnamefont {G.}~\bibnamefont
  {Martinez-Pinedo}},\ }\href {\doibase
  https://doi.org/10.1016/S0375-9474(00)00131-7} {\bibfield  {journal}
  {\bibinfo  {journal} {Nuclear Physics A}\ }\textbf {\bibinfo {volume}
  {673}},\ \bibinfo {pages} {481 } (\bibinfo {year} {2000})}\BibitemShut
  {NoStop}%
\end{thebibliography}%

\end{document}